\documentclass[aps,pra,twocolumn,showpacs,preprintnumbers]{revtex4-1}

\usepackage{psfrag,graphicx}
\usepackage{dcolumn}
\usepackage{amsmath,amssymb}
\usepackage{bm}
\usepackage{amsfonts,amssymb,amsmath}        
\usepackage{epstopdf}
\usepackage{amsthm}
\usepackage{hyperref}
\usepackage{color}

\newcommand{\be}{\begin{equation}}
\newcommand{\ee}{\end{equation}}
\newcommand{\bq}{\begin{eqnarray}}
\newcommand{\eq}{\end{eqnarray}}

\newcommand{\ket}[1]{\left | \, #1 \right\rangle}
\newcommand{\bra}[1]{\left \langle #1 \, \right |}

\newcommand{\ZZ}{\mathbb{Z}}

\begin{document}

\title{Improved HDRG decoders for qudit and non-Abelian quantum error correction}
\author{Adrian Hutter, Daniel Loss, James R. Wootton}
\affiliation{Department of Physics, University of Basel, Klingelbergstrasse 82, CH-4056 Basel, Switzerland}

\date{\today}

\begin{abstract}
Hard-decision renormalization group (HDRG) decoders are an important class of decoding algorithms for topological quantum error correction.
Due to their versatility, they have been used to decode systems with fractal logical operators, color codes, qudit topological codes, and non-Abelian systems.
In this work, we develop a method of performing HDRG decoding which combines strenghts of existing decoders and further improves upon them.
In particular, we increase the minimal number of errors necessary for a logical error in a system of linear size $L$ from $\Theta(L^{2/3})$ to $\Omega(L^{1-\epsilon})$ for any $\epsilon>0$.
We apply our algorithm to decoding $D(\ZZ_d)$ quantum double models and a non-Abelian anyon model with Fibonacci-like fusion rules, and show that it indeed significantly outperforms previous HDRG decoders.
Furthermore, we provide the first study of continuous error correction with imperfect syndrome measurements for the $D(\ZZ_d)$ quantum double models.
The parallelized runtime of our algorithm is $\text{poly}(\log L)$ for the perfect measurement case.
In the continuous case with imperfect syndrome measurements, the averaged runtime is $O(1)$ for Abelian systems, while continuous error correction for non-Abelian anyons stays an open problem.
\end{abstract}

\maketitle

\section{Introduction}

Over the last decade, topological error correcting codes have emerged as the primary candidate for quantum error correction \cite{kitaev,dennis}. Errors in these codes can be interpreted in terms of the creation, transport and annihilation of quasiparticles, allowing the design of intuitive decoding algorithms \cite{fowler_fast,sdrg,hutterwootton,dennis_thesis,expand}. The anyonic nature of the quasiparticles also makes them well suited to implement quantum computation on the stored information \cite{fowler_rev,wootton_rev}.

Recently a novel class of decoding algorithms was introduced for topological quantum error correcting codes \cite{bravyihaah,abcb}. They were prominently used for correcting codes with fractal logical operators \cite{haah}, for which no alternative decoding procedure was available. These decoders have since been referred to as `hard-decision renormalization group' or `HDRG' decoders \cite{raus_hdrg}.

The main advantage of HDRG decoders arises when they are applied to codes for which syndrome measurements do not have a simple binary output, but instead give more detailed information. Properly taking this information into account will greatly improve the success rate of a decoding algorithm, but will also greatly increase the run-time. The design of HDRG decoders allows them to make a compromise, providing decoding that is fast but successful.

These decoders are also hugely important to the emerging field of non-Abelian decoding \cite{wootton_na,brell_na}. For these much of the additional syndrome information is not initially accessible. The method by which it can be extracted (fusing anyons and observing the fusion outcome) exactly mirrors the way in which it is used within HDRG decoders. Their development is therefore vitally important for topological quantum computation.

Finally, HDRG decoders are also relevant for correcting finite-temperature quantum memories \cite{memories_review}, a purpose for which they have been employed in Refs.~\cite{bravyihaah,brown}. A quantum memory model of particular recent interest, for which decoding is an open problem and for which HDRG methods might prove useful, is developed in Ref.~\cite{brell}.

HDRG decoding was first introduced in Ref.~\cite{harrington_thesis}. Based on ideas from Ref.~\cite{harrington_thesis}, Ref.~\cite{bravyihaah} developed an HDRG decoder that was designed to be generally applicable to topological codes, and also to allow an analytic proof that it realizes a finite threshold error rate for local noise. 
However, it was later shown that developments to the method can allow better decoding \cite{abcb}. Here we expand upon this work. We consider strengths and weaknesses of the existing methods, and determine how the strenghts of the different decoders can be combined and how they can be improved further.
In particular, we increase the minimal number of errors necessary for a logical error in a code of linear size $L$ from $\Theta(L^{2/3})$ to $\Omega(L^{1-\epsilon})$ for any $\epsilon>0$.

For concreteness we consider a particular choice of topological codes to act as a sandbox, namely the $D(\ZZ_d)$ quantum double models \cite{kitaev}, the qudit generalization of the more familiar qubit toric code. 
However, our results will apply more generally to other types of anyonic systems.
Systems with qudits of internal dimension higher than $2$ are of interest for quantum computing due to the possibility of magic state distillation with improved error thresholds and reduced overhead \cite{ACB,CAB} and of transverse non-Clifford gates \cite{earl}. The possibility of implementing quantum computation with these codes was explored in Ref.~\cite{wootton_qudit}.

We consider the case of perfect syndrome measurements, which has been studied previously using both HDRG and non-HDRG decoders \cite{sdrg,abcb}. We also do the first study of these codes for imperfect syndrome measurements, which we model using measurement outcome errors. Finally, we employ the developed methods for decoding the non-Abelian $\Phi$-$\Lambda$ model. We find for this model a threshold error rate of $15\%$, while previous HDRG methods achieved $7\%$. 

The rest of this paper is organized as follows.
Sec.~\ref{sec:DZd} briefly introduces the $D(\ZZ_d)$ quantum double models, which serve as a testbed in the following sections.
Sec.~\ref{sec:HDRG} defines HDRG decoders and introduces decoders used in the previous literature.
Sec.~\ref{sec:improving} discusses strengths and weaknesses of different decoders and how they can be improved upon.
In Sec.~\ref{sec:MWPM} we present a minimum-weight perfect matching based HDRG decoder, which incorporates the lessons learned in Sec.~\ref{sec:improving}.
We apply our decoder to the $D(\ZZ_d)$ model in Sec.~\ref{sec:results} and to a non-Abelian anyon model in Sec.~\ref{sec:nonabelian}.
We discuss the run-time of our algorithm in Sec.~\ref{sec:runtime} and conclude in Sec.~\ref{sec:conclusions}.

\section{$D(\ZZ_d)$ quantum double models}\label{sec:DZd}

\begin{figure}
\centering
\includegraphics[width=0.8\columnwidth]{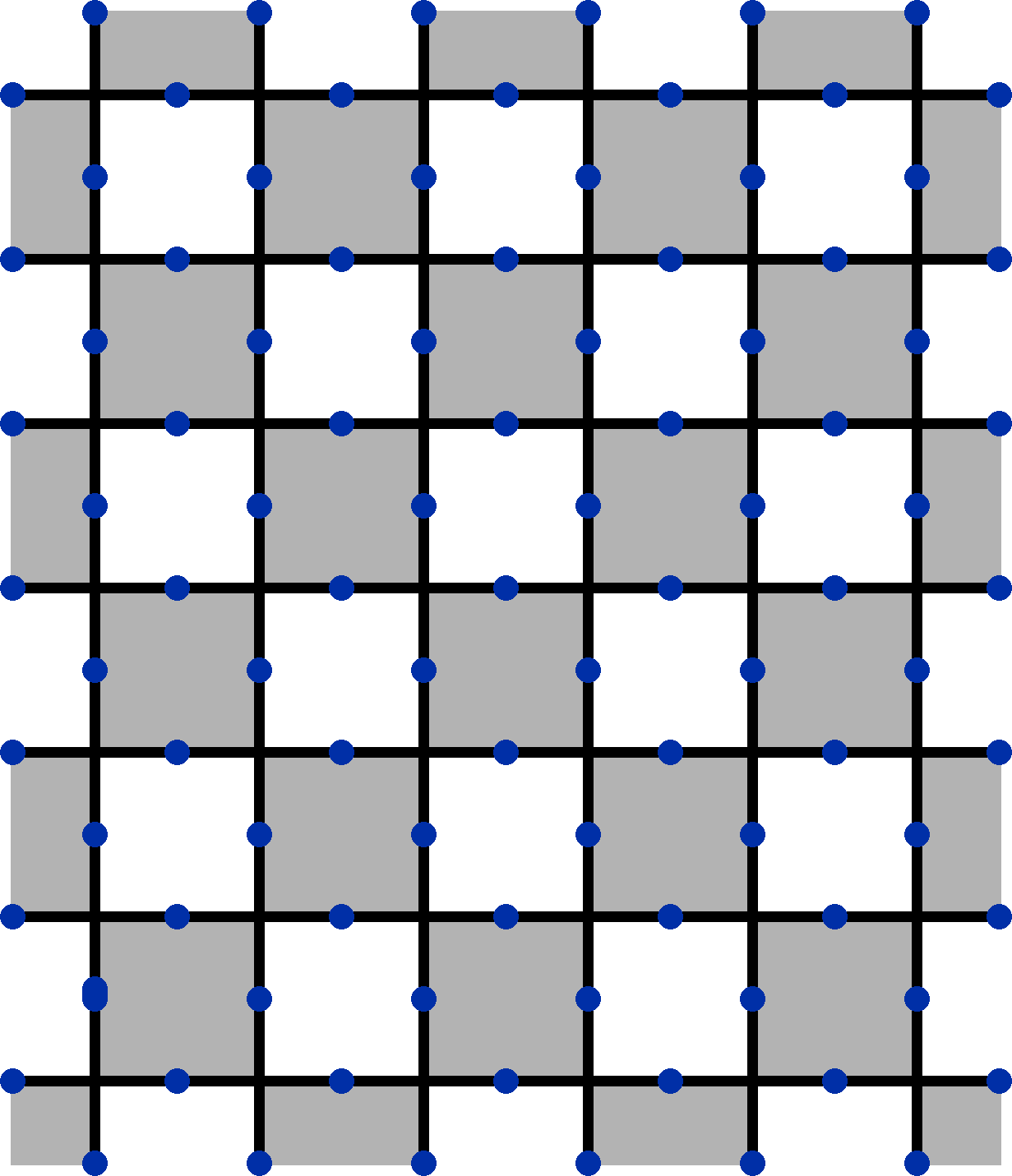}
\caption{Spin lattice on which the codes are defined, with spins placed on edges.}
\label{fig:lattice}
\end{figure}

First we introduce the topological error correcting codes on which the methods we develop will be tested: the $D(\ZZ_d)$ quantum double models \cite{kitaev}. In particular we consider their planar variant, defined on the spin lattice shown in Fig. \ref{fig:lattice}.

Stabilizer operators for these codes are defined on the qudits around the plaquettes and vertices of the lattice. The plaquette and vertex operators are independent of each other, and also dual to each other. We can thus consider only the plaquette operators for simplicity, since all results will apply to the vertex operators also.
For a more detailed introduction, the reader is referred to Ref.~\cite{bullock}, which provides the first study of $\ZZ_d$ gauge theories as error correcting codes.

To define the plaquette operators we bicolour the plaquettes black and white in chessboard fashion. On white plaquettes these stabilizers are defined as 
\be
B_p = \prod_{j \in p} \sigma^z_j.
\ee
Here the product is over each qudit $j$ around the plaquette $p$. The $\sigma^z$ operator is a qudit generalization of the standard Pauli operator. This is defined as 
\be
\sigma^z = \sum_{j=0}^{d-1} e^{i \omega j} \ket{j}\bra{j}, \,\,\,\, \omega = \frac{2 \pi}{d},
\ee 
for a $d$-level qudit. 
The plaquette operators for black plaquettes are simply defined as $B_p^\dagger$.

The plaquette operators have $d$ possible eigenvalues. These correspond to the $d$th roots of unity $\omega^g$ for $g=0, \ldots, d-1$. Syndrome measurements determine the value of $g$ for each plaquette. The case of $g=0$ is the trivial syndrome, and is associated with anyonic vacuum, $1$ on the corresponding plaquette. All other values of $g$ correspond to unique anyon types $m_g$. The value $g$ is referred to as the magnetic charge, or simply the charge, of the anyon.

The syndrome is affected by single spin operators of the form
\be
(\sigma^x)^g = \sum_{j=0}^{d-1} \ket{j+g\,\,\text{ mod }\,\,d}\bra{j}.
\ee
The effects of these on a spin will be to create an anyon of type $m_g$ in the white plaquette adjacent to the qudit on which it was applied, and one of type $m_{d-g}$ in the black plaquette. If anyons are already present on these plaquettes they will fuse with the newly created ones according to the fusion rules
\be
m_g \times m_h = m_{g + h\,\,\text{ mod }\,\,d}.
\ee
Here $m_0 = 1$. Note that the antiparticle of any $m_{g}$ is $m_{d-g}$. Henceforth we will refer to the latter simply as $m_{-g}$.

Given these operations it is possible to move anyons. An anyon of type $m_g$ on a white black can be moved onto a neighbouring black plaquette by applying $(\sigma^x)^{-g}$ to the qudit between them. This creates an $m_{-g}$ anyon in the white plaquette and an $m_g$ on the black. The former annihilates the original anyon, and so results in its effective movement to the black plaquette. Corresponding operators can be applied for other cases.

Given this means of transport, the minimum number of qudits on which these operations must be applied in order to move an anyon from one plaquette to another is the Manhattan distance ($L_1$ metric) between them. It is therefore this metric that we use to evaluate distances between anyons.

The stabilizer space of the code is defined as that for which all plaquettes and vertices hold vacuum. This space is $d^2$ dimensional, and so capable of storing two logical qudits. The effect of errors acting on a state initially in the stabilizer space is to create anyons, and then to move, split and fuse them. The pattern of errors applied in any given case is called the error chain, $E$. The resulting pattern of anyons is the syndrome, $S$.

The job of a decoding algorithm is to remove the effects of the errors. It must therefore remove the anyons by annihilating them with each other. In principle this would be done by applying operators of the form $(\sigma^x)^h$ to the spins. However, this is unnecessary in practice. Instead the operations can be performed effectively by accounting for them in all future measurements and operations on the effected spins. The total operation applied (either actually or virtually) is known as the recovery operator, $R$. The error correction is successful if the total effect of errors and correction, $RE$, is a product of the stabilizer operators. This is satisfied as long as $RE$ contains no loops of errors that wrap around the non-trivial cycles of the torus.

We consider a simple error model that has previously been used to benchmark decoders for this code. This is that of $(\sigma^x)^g$ type errors applied independently to each physical qubit. The strength of the noise is parameterized $p$, which denotes the probability for each qudit that an error of this form with $g \neq 0$ is applied. We consider the case that all non-zero $g$ are applied with equal probability $p/(d-1)$.

\section{HDRG decoders}\label{sec:HDRG}

Until now only the decoder of Ref.~\cite{bravyihaah} and its derivatives have been referred to as HDRG in the context of topological codes. However, in this work we use the term to refer to a more general class of decoders.

In order to define this class, we must first introduce some terminology. Subsets of the syndrome, $S$, are referred to as clusters. A cluster is said to be neutral if it is possible for it to be removed without otherwise affecting the syndrome. Otherwise the cluster is non-neutral. For the $D(\ZZ_d)$ codes a cluster, which is a set of anyons, is neutral if the sum of their charges is zero modulo $d$. A set of errors that creates a single neutral set of anyons is called an error net. Those that create only two anyons at their endpoints are known as an error string.

The class of decoders we consider are those that use the repeated application of the following process. Initially, each non-trivial element of the syndrome is considered to be a separate cluster.
\begin{enumerate}
\item Form at least one new cluster by combining existing clusters.
\item Check for each new cluster whether it is neutral, and find a neutralization operator $R_j$ for each neutral cluster $C_j$.
\item Update $S$ by removing all neutral clusters.
\end{enumerate}
This continues until the syndrome is empty. The decoder then outputs $R = \prod_j R_j$ as a proposed correction operator.

Note that once elements of the syndrome are included in the same cluster, they will remain within the same cluster for the rest of the process. It is this feature that allows the procedure to be applicable to non-Abelian anyons, since in that case neutrality tests are performed by the irreversible act of fusion.

Only the first step of this process is not uniquely defined. The exact means by which the clustering is performed is what distinguishes the different HDRG decoders. Below we present the HDRG decoders that have been applied to topological codes so far.

\subsection{BH and ABCB decoders}

The HDRG decoders of Refs.~\cite{bravyihaah} (BH) and \cite{abcb} (ABCB) work as follows. Firstly they define a physical distance $d_{j,k}$ between all pairs of non-trivial syndrome elements $j$ and $k$. For BH the Chebyshev distance ($L_{\infty}$ metric) is used, whereas for ABCB this is a combination of the Chebyshev distance and Manhattan distance ($L_1$ metric). A search distance $D(n)$ is also defined for the $n$th iteration of the algorithm. For BH $D(n)=2^n$, whereas for ABCB it is simply $D(n)=n+1$. The algorithm then runs through the following steps.
\begin{enumerate}
\item Form a graph with a vertex corresponding to each non-trivial syndrome element and no edges. Set $n=0$.
\item Add an edge between all pairs of vertices for which $d_{jk} \leq D(n)$.
\item Clusters are connected components of this graph. Check all clusters for neutrality. Remove all vertices corresponding to each neutral cluster $C_j$.
\item If vertices remain, increment $n$ by $1$ and repeat from step 2. Otherwise proceed to step 5.
\item For each neutral cluster $C_j$ find an operator $R_j$ that acts only on the spins in its neighbourhood, the action of which would remove the syndrome.
\item Output the total recovery operator $R = \prod_j R_j$.
\end{enumerate}

An `enhanced' version of the ABCB decoder has also been considered Ref.~\cite{abcb}. This has an initialization step in which neutral clusters are searched for over a small area. The search is performed such that elements of the syndrome included within the same cluster at one point included within different clusters later. This enhancement is therefore no longer an HDRG decoder according to our definition.

\subsection{Expanding diamonds decoder}

We consider the variant of the expanding diamonds algorithm \cite{dennis_thesis,expand} presented in \cite{expand}.
This also requires distances $d_{jk}$ and $D(n)$, with the Manhattan distance used for the former and $D(n)=n+1$ for the latter. The clustering at the $(n+1)$th iteration is done by finding pairs of mutually nearest neighbouring clusters in the $n$th iteration. It does this as follows.
\begin{enumerate}
\item Assign each non-trivial syndrome element its own cluster, and label these from $1$ to $N_0$ (the number of non-trivial syndrome elements). Set $n=0$.
\item Number the clusters left to right and top to bottom. Loop through them in this order. For each cluster, $j$, check whether there exists a cluster $k>j$ for which $d_{jk}<D(n)$. If so, merge the clusters. If any such cluster is neutral, remove it from the syndrome. 
\item Label the $N_{n+1}$ clusters that remain from $1$ to $N_{n+1}$. Set the distance $d_{jk}$ between clusters $j$ and $k$ to be the minimum distance from an anyon of one to an anyon of the other.
\item For $N_{n+1}>0$, increment $n$ by $1$ and repeat from step 2. Otherwise proceed to step 6.
\item For each neutral cluster $C_j$ find an operator $R_j$ that acts only on the spins in its neighbourhood, the action of which would remove the syndrome.
\item Output the total recovery operator $R = \prod_j R_j$.
\end{enumerate}

\section{Improving HDRG decoders}\label{sec:improving}

One major difference between the algorithms described above is the speed at which they increase cluster size. Expanding diamonds does this very slowly, with each new cluster formed out of only two previous ones. The BH and ABCB decoders do it more quickly, with the exponentially increasing search distance of BH making it the fastest of all.

It is natural to ask which speed of cluster increase leads to the best results. Both extremes have their advantages. Slow increase of cluster size means that the clusters checked for neutrality will typically contain less anyons. This therefore reveals more information about their relative charges. When clusters are typically large, this information is far more coarse grained.

Smaller cluster size also means that there will be more clusters, and hence more neutrality checks. Although this may seem like an advantage, recall that any cluster found to be neutral will be removed from the syndrome in all of the HDRG decoders above.

If the resulting annihilation operator for the anyons within the cluster is topologically equivalent to the error that created them, this removal poses no problems. However, this may not be the case. Consider a cluster composed of two anyons, one of type $m_a$ and one $m_{-a}$. Since these are antiparticles, they could have been created by a single error string. However, it is also possible that they were created by different error strings, whose other endpoints lie outside the cluster. The fact the cluster is neutral is then due only to random chance, and does not correspond to successful correction from the decoder. Discarding information about these neutral clusters makes it impossible for the decoder to realize and correct its mistakes. This therefore can give an advantage to algorithms with quickly growing cluster size, since they are more careful about declaring clusters neutral.

In summary, slowly increasing clusters lead to more syndrome information being extracted and used by the decoder. However, it also leads to more being lost as neutral clusters are found. Quickly increasing clusters extract less of the syndrome, but also lose less. It is not clear which speed of cluster increase leads to maximal syndrome usage, and so which should lead to best decoding.

Rather than searching for the optimal speed, we will consider how the advantages might be combined and the disadvantages negated. This can be achieved using an algorithm with slowly increasing cluster size, but which does not completely forget about the neutral clusters. The challenge then is to determine how information about neutral clusters might be used in a way that does not affect the efficiency of HDRG decoders.

The simplest way to carry forward information about neutral clusters is by using a simple modification of the physical distance. To motivate this, consider two strings of errors along a line. Each are length $l_0$, and create an anyon of type $m_a$ on their left and $m_{-a}$ on their right. The distance between the two strings is $l_0-1$. Both expanding diamonds and ABCB would see that the shortest distance between two anyons is that between the $-a$ of the left string and the $a$ of the right. They would then form a cluster out of these, see it is neutral and remove it from the syndrome. The same is true of BH if $l_0$ is a power of two. However, we will restrict our attention to the other decoders for simplicity.

This action taken by the decoders is a mistake. However, this mistake will not lead to any ill effects as long as the remaining $m_a$ from the left and $m_{-a}$ from the right end up in the same cluster (without looping around the torus). This will certainly happen if no anyon is closer to either than the other. However, note that the distance between them is $3l_0-1$. This does not just include the $2l_0$ errors that occurred between them, but also the $l_0-1$ gap. The distance between the anyons should really only reflect the number of errors required to connect them. This increased distance makes them less likely to find each other than they should be.

This issue can be solved by recalling the existence of the neutral cluster. The number of errors required to connect the two anyons is only that needed to connect them both to the neutral cluster, and so the distance should be defined accordingly. This would then give the correct distance $2l_0$ between them. Whenever a neutral cluster $C = \{c_1, c_2, \ldots\}$ is found, the physical distance between the remaining clusters should thus be updated according to
\be
d_{jk} \rightarrow \min \left( d_{jk}, \,\,\, \min_{c_m, c_n \in C} \left( d_{jc_m} + d_{c_mk} \right) \right).
\ee
By allowing the distance to take shortcuts between neutral clusters, information about their positions is retained by the decoder. Also note that this principle is not restricted to neutral clusters, and so shortcuts via non-neutral clusters can also be used.

\subsection{Example: Cantor-like error chains}\label{sec:cantor}

The effectiveness of the redefined physical distance can be seen by considering Cantor-like error chains \cite{dennis_thesis,expand}. These can cause all of the decoders considered above to fail with only $\Theta(L^\beta)$ errors, where $\beta<1$, when the shortcuts are not used. The use of the shortcuts, however, means that the required number of errors is asymptotically greater than $\Theta(L^\beta)$ for any $\beta<1$ (though not as high as $\Theta(L)$).
The minimal number of errors that make a decoder fail is of practical relevance since the failure rate of the decoder is exponentially suppressed with the corresponding exponent in the low-$p$ limit. 

Consider again the two error strings discussed above, which both have an anyon of type $a$ on their left and $-a$ on their right. They are both of length $l_0$ and lie along a line. We will use $g_0$ to denote the distance between them, and we will refer to any such pair of strings as a level-$1$ bundle. Note that the total length of a level-$1$ bundle, including the gap, is $l_1 = 2 l_0 + g_0$.

We similarly define a level-$(n+1)$ bundle to be a pair of level-$n$ bundles along the same line and with a gap $g_n$ between them. The size of a level-$(n+1)$ bundle is then $l_{n+1} = 2 l_n + g_n$.

Let us consider the case of a level-$m$ bundle such that $l_m \geq \lfloor(L+1)/2\rfloor$. If $g_0$ is significantly smaller than $l_0$, the decoders will incorrectly annihilate the inner two anyons of each level-$1$ bundle. Each level-$2$ bundle will then be composed of two strings of length $l_1$ with a gap of $g_1$ between them. Again, $g_1$ being significantly smaller than $l_1$ will lead to incorrect annihilation. If all $g_n$ are significantly smaller than the corresponding $l_n$, this chain of mistakes will lead to a pair of anyons separated by $l_m \geq \lfloor(L+1)/2\rfloor$. This will then lead to a logical error (with probability $\frac{1}{2}$ if $L$ is even and $l_m=L/2$, and with certainty in all other cases).

The exact requirements for $g_n$ and $l_n$ required to cause a logical error depend on the decoders. We wish to consider fatal error chains with the smallest number of errors, and so the largest possible gaps. For the expanding diamonds and ABCB decoders, a logical error will occur when $g_n < l_n$ $\forall n$. We will therefore consider the minimal case of $g_n = l_n-1$. For simplicity we will also use $l_0=2$.

In this case, the length of a level-$n$ bundle will follow
\be
l_{n+1} = 2 l_n + g_n = 3 l_n -1 = \frac{3^{n+1}+1}{2}\,.
\ee
A level-$m$ bundle with $l_m\geq\lfloor(L+1)/2\rfloor$ then requires $m \geq \lceil\log_3 (L-1)\rceil$. The number of errors within any level-$n$ bundle is clearly $2^{n+1}$. The total number of errors required to cause a logical error is then $\Omega(L^\beta)$, where $\beta = \log_3 2\approx0.63$.

For BH, the corresponding minimal condition for a logical error is
\be
g_n = 2^{\left \lceil \log_2 l_n \right \rceil -1 }, \,\, \forall n\,.
\ee
This reflects the fact that the search distance $D(k)$ treats all distances from $2^{k-1}+1$ to $2^k$ the same for any $k$. The length of a level-$n$ bundle is then
\be
l_{n+1} = 2 l_n + 2^{\left \lceil \log_2 l_n \right\rceil  - 1}\,.
\ee
Assume that $l_n=2^k+c$ with $0<c\leq2^{k-1}$. For any $l_0$, either $l_0$ or $l_1$ is of this form. Then, $l_{n+1}=3\times2^k+2c$ and $l_{n+2}=2^{k+3}+4c$. Note that the first summand grows by a factor of $8$ while the second summand grows by a factor of $4$, such that the latter becomes vanishing relative to the former. So asymptotically, the ratio $l_{n+1}/l_n$ oscillates between $3$ and $\frac{8}{3}$, and hence $l_n=(2\sqrt{2})^{n+O(1)}$. A level-$n$ bundle with $l_n\geq\lfloor(L+1)/2\rfloor$ then requires $n\geq\log(L)/\log(2\sqrt{2})+O(1)$ and thus involves at least $2^{n+1}=\Theta(L^\beta)$ errors with $\beta=\frac{2}{3}$. The exponent $\beta=\frac{2}{3}\approx0.67$ is a slight improvement over expanding diamonds and ABCB, but not greatly so.

When the redefined distances are used, the error chains considered above will no longer lead to logical errors. Instead let us define the width of a bundle to be the distance between its extremal anyons when all others have been annihilated incorrectly. Taking the shortcuts into account, this obeys $w_{n} = 2^n l_0$. Note that $w_n$ is then equal to the number of errors in a level-$n$ bundle.

For expanding diamonds and ABCB the requirement for a logical error is now $g_n < w_n$. The total length of a  minimal bundle leading to a logical error (i.e., $g_n=w_n-1$) then obeys
\begin{align}
l_{n+1} &= 2 l_n + g_n \nonumber\\ &= 2 l_n + 2^n l_0 -1 \nonumber\\ &= (n+1)2^n+1 \nonumber\\ &= \Theta(n 2^n)\,.
\end{align}
For BH the corresponding condition for a logical error is
\be
g_n = 2^{\left \lceil \log_2 w_n \right \rceil -1 }, \,\, \forall n\,.
\ee
Considering again the case of $l_0=2$ gives $g_n = 2^n$, leading to $l_{n+1} = \Theta(n 2^n)$.

All of the decoders considered therefore result in the same scaling $l_{n+1} = \Theta(n 2^n)$ for minimal uncorrectable error chains when the shortcuts are used. 
We thus have $l_n = O( (2+ \epsilon)^n)$ for any $\epsilon>0$. 
In order to create a logical error, we need $l_m\geq\lfloor(L+1)/2\rfloor$ and therefore a bundle of level $n=\Omega(\log_{2+\epsilon}L)$, which involves $w_n=2^nl_0=\Omega(L^{\log_{2+\epsilon} 2})$ errors.
This is higher than any $\Theta(L^\beta)$ for $\beta<1$, but does not reach the value of $\beta=1$ that non-HDRG decoders may realize. Nevertheless it is a marked improvement over $\beta=\log_3 2 \approx 0.63$ and $\beta =2/3\approx 0.67$.

Note that using the shortcuts, the smallest code which can lead to a logical error with less than $\lfloor(L+1)/2\rfloor$ errors is of size $L=9$. For such a code, a level-$1$ bundle leads to a logical error with probability $\frac{1}{2}$.

\section{Minimum Weight Matching HDRG decoder}\label{sec:MWPM}

We now introduce a novel decoder, based on the lessons learned above. Like expanding diamonds, this will have a slow increase of cluster size for which each new cluster will be composed of two previous ones. However, the means by which the clustering is performed will not be based on a search distance. Instead it will use a generalization of the minimum weight perfect matching algorithm that gives high quality decoding in the $D(\ZZ_2)$ case \cite{fowler_fast}. Shortcuts will also be used.

\subsection{Minimum Weight Matching Algorithm}

The backbone of the decoder is an algorithm for finding the minimum weight matching (MWM) of a graph. This is in turn based upon an algorithm for minimum weight perfect matching (MWPM).

A perfect matching is a decomposition of the vertices of a graph into pairs. This must be such that the two vertices, $j$ and $k$, of each pair are connected by an edge $jk$ of the graph. For a weighted graph, each edge $jk$ will have a weight $W_{jk}$. We can then associate a total weight to a perfect matching by summing the weights of the edge corresponding to each pair. A minimum weight perfect matching is such a pairing that achieves minimal weight. Clearly, a MWPM can only exist for graphs with an even number of vertices.

A non-perfect matching does not cover all vertices. It corresponds to a partial pairing of the vertices, with some vertices left unpaired. In order to define a total weight for a such a matching, let us assign a weight $W_j$ to each vertex $j$. All paired vertices then contribute their corresponding edge weight to the total, and all unpaired vertices contribute their vertex weight.

Any algorithm that is able to find minimum weight perfect matchings of graphs will also be able to find minimum weight matchings according to this definition. To do to this for a weighted graph $G$ we create a graph $G'$. This includes two vertices $j$ and $j'$, for each vertex $j$ of $G$. Every edge $jk$ in $G$ corresponds to edges $jk$ and $j'k'$ in $G'$ with weights
\be
W'_{jk} = W_{jk}, \,\,\, W'_{j'k'} = 0\,.
\ee
The graph $G'$ also includes edges $jj'$ for each $j$ of $G$. The weight of these is set to the vertex weight: $W_{jj'} = W_j$.

For the graph $G'$ constructed in this way, a pair can only take three forms: $jk$, $j'k'$ and $jj'$. The $jk$ type pairs correspond to a pairing in the graph $G$ and has corresponding weight $W_{jk}$. For each of these a corresponding $j'k'$ pair can occur in order to ensure that the matching is perfect with zero weight. The pairs of the form $jj'$ correspond to a vertex of $G$ that does not pair with anything, and have the corresponding weight $W_j$. Any MWPM of $G'$ therefore corresponds directly to a MWM of $G$.

Algorithms to efficiently find the MWPM of a graph are well known \cite{blossom,fowler_fast}. These can therefore be used to implement the following decoding method.

\subsection{Decoding algorithm}

Each anyon of the syndrome is associated with the vertices of a graph, $G$. In general we will consider this to be a complete graph, with an edge between each pair of vertices. However, not all edges will need to be considered in practice.

Each edge is assigned a weight whose value depends on the distance between the corresponding anyons. Each vertex is assigned a weight that depends on the distance from the anyon to its nearest neighbours. These weights are defined in more detail in the following sections.

Given the weighted graph $G$, the MWM algorithm is run in order to find a set of non-overlapping anyon pairs. These pairs are treated as clusters, and are therefore checked for neutrality.

For each non-neutral pair, the corresponding vertices $j$ and $k$ are combined into a single vertex $(jk)$. The edge between $j$ and $k$ is removed. The edge weights and vertex weights are refined for the new cluster as explained in the following sections.

For each neutral pair the corresponding vertices are removed from the graph, as are all edges incident upon them. Since all weights are based on the distances between anyons, the weights for remaining edges and vertices should be updated in order to take advantage of shortcuts via the neutral cluster. Shorcuts via non-neutral clusters are also considered.

This process is repeated on the resulting graph until all vertices have been removed. The final recovery operation is the product of annihilation operators for each neutral cluster.

\subsection{Pairing Weight}

Consider a specific error chain, $E$, which contains $|E|$ errors. The probability of this, up to normalization, is
\be
P(E) = \left( \frac{p/(d-1)}{1-p} \right)^{|E|} = e^{-\beta |E|}\,.
\ee
Here $\beta$ is defined as
\be\label{eq:beta}
\beta = - \log \left( \frac{p/(d-1)}{1-p} \right)\,.
\ee
In order to motivate the definition of the pairing weights $W_{jk}$, let us consider a modified error model. This acts according to the standard error model defined above, except that no splittings or fusions are allowed. All error nets are therefore strings: they simply create two anyons that are the antiparticles of each other. Since there are $d-1$ types of different non-trivial particle, there are $d-1$ types of string.

For this case, one possible tactic for an HDRG decoder is to consider all possible error chains $E$ that are consistent with the syndrome and determine which is most likely. The resulting pairs of anyons are then used as the clusters.

The most likely error strings are those that have the smallest number of errors. For each pair of anyons, $j$ and $k$, created by the same error string, the minimum number of errors is $d_{jk}$. The probability of the corresponding error chain can then be expressed as
\be
P(E) = \prod_{(j,k)} e^{-\beta d_{jk}}\,.
\ee
Note that this probability assumes that the path and type of error string between each pair is specified. However, the decoder does not care about this information. It wants to find the most probable pairing, without regard to the path that the errors took between the anyons. Also, since the decoder is HDRG, it does not use the anyon charge information when performing the clustering. It therefore does not care which of the $d-1$ possible types of error string occurred in each case.

Let us use $\{E\}$ to denote the set of all error chains with the same pairing as $E$, that differ only in path and type of the error string. Let us also use $\mu_{j,k}$ to denote the number of minimum distance error strings between $j$ and $k$, including the multiplicities in both path and error type,
\be
\mu_{jk} = (d-1) {d_{jk} \choose d^x_{jk}}\,.
\ee
Here $d^x_{jk}$ denotes the distance between $j$ and $k$ in the $x$ direction, such that the Manhattan distance can be expressed $d_{jk} = d^x_{jk} + d^y_{jk}$. The probability for the set $\{E\}$ is then
\be
P(\{E\}) = \prod_{(j,k)} \mu_{jk} \,\, e^{-\beta d_{jk}}\,.
\ee
The task of finding a pairing that maximizes $P(\{E\})$ is clearly equivalent to one that minimizes $-\log P(\{E\})$. It can thus be achieved using MWPM using the following weight for each pair
\be \label{pair_weight}
W_{jk} = d_{j,k} - (\log \mu_{j,k}) / \beta\,.
\ee
Even though these weights are defined for an alternative error model, and we will need to use MWM rather than MWPM for the true error model, we will continue to use these weights. The vertex weights will then be defined such that the whole minimization problem is consistent with the true error model.

When two anyons (or non-neutral clusters), $j$ and $k$, are combined to form a non-neutral cluster $(jk)$, the weight for this cluster to be paired with another anyon or non-neutral cluster $l$ must be defined. This is done by defining the distance between $(jk)$ and $l$ to be
\be
d_{(jk)l} = \min ( d_{jl}, d_{kl} )\,.
\ee
The multiplicity $\mu_{(jk)l}$ is taken to be $\mu_{jl}$ if $d_{jl}<d_{kl}$, $\mu_{kl}$ if $d_{kl}<d_{jl}$, and  $\mu_{jl}+\mu_{kl}$ if the distances are equal.

The distances and multiplicities must also be modified to take shortcuts into account, via both neutral and non-neutral clusters. For clusters $j$ and $k$ connected via a cluster $l$ the distance becomes
\be
d_{jk} \rightarrow \min ( d_{jk}, d_{j,l} + d_{l,k} )\,.
\ee
If the latter distance via the cluster $k$ is indeed minimal, the multiplicity is updated according to
\be\label{eq:multUpdate}
\mu_{jk} = \mu_{jl} \,\, \mu_{lk}\,,
\ee
while this expression is added to $\mu_{jk}$ if the two distances are equal.
Note that this introduces an extra factor of $d-1$ for every cluster that the shortcut goes via. This would be expected for non-neutral clusters, since the anyon deposited by the error string from $j$ does not need to have any relation to that deposited by the string from $k$. However these should be antiparticles for neutral clusters, and this restriction should mean that this extra factor is not included. However, for simplicity we use Eq.~(\ref{eq:multUpdate}) irrespective of the anyonic charge of the cluster.

These methods of updating the distances and multiplicities for the edge weights also apply to their use within the vertex weights, as defined below.

\subsection{Tag-along weight}

The true error model does include splittings and fusions. Therefore, the most likely error chain will not typically be composed only of strings, but more general error nets. However in order to motivate our choice of the vertex weights $W_j$ in the graph $G'$, we will again consider a restricted error model, allowing only error nets composed of strings that meet at anyons. Like the pairing, this also allows us to associate error chains with edge covers.

It is clear that the edge cover corresponding to the most likely error chain will not contain simple cycles. This is because edges can be removed from these (and hence the probability will increase) while maintaining the edge cover. All disconnected subgraphs will therefore be trees. This same argument can be applied to any tree that is not a star graph. A star that contains $n$ vertices has $n-1$ external vertices which are incident upon only one edge and one internal vertex incident upon $n-1$ edges. A pair, for $n=2$, is the simplest example of this.

A moment's thought shows that the most likely error chain contains only stars which are either of size $2$ or for which the internal vertex is each externel vertex' nearest neighbor. To see this, assume by contradiction that the nearest neighbor of an external vertex in a star of size larger than $2$ is not the internal vertex. It is thus either another external vertex of the same star or an internal or external vertex of another star. In each of these cases, we can connect the external vertex to its nearest neighbor, remove the edge connecting it to the internal vertex, and potentially remove further edges as well. It is thus always possible to decrease the weight and increase the likelihood of such an error net.

Let us define two of the vertices from each star, one internal and one external, to be a pair. All other external vertices are defined to be `tag-alongs' to that pair.

The MWM algorithm can then be used to decompose the anyons into pairs and tag-alongs. The tag-alongs are those anyons that are not paired by the algorithm. They can be considered to be tagging-along with any of their nearest neighbours. 

We use 
\be
d_j = \min_{k} d_{j,k}\,.
\ee
to denote the nearest neighbour distance of an anyon.
The weight that MWM assigns to each pair will be the pairing weight of Eq. (\ref{pair_weight}). For the tag-along weight, note that the decoder only combines the two anyons (or non-neutral clusters) within each pair to form new clusters. The tag-alongs are not included. It is thus not the most likely decomposition of the errors into stars that is most important, but the most likely decomposition into pairs and tag-alongs. The tag-along weight should therefore incorporate information about the number of nearest neighbours it can tag-along with. To do this we define the tag-along multiplicity of an anyon to be
\be
\mu_j = \sum_{k \in {\rm nn}(j)} \mu_{j,k}\,.
\ee
Here ${\rm nn}(j)$ denotes the nearest neighbours of $j$, and so $\mu_j$ is the sum of all possible minimum distance error strings to nearest neighbours. 
The tag-along weight is then defined as
\be 
W^T_j = d_{j} - (\log \mu_{j})/\beta\,,
\ee
for each anyon, $j$.

\subsection{Abstaining weight}

The true error model does not restrict to the error nets considered above, where strings meet only at anyons. Instead it can have more general structures, such as a triskelion with an anyon at each foot. Such error nets can cover the anyons using less errors than when only strings meeting at anyons are considered. The tag-along weights considered above are thus often an overestimate.

In order to deal with this, we will consider an alternative definition of the vertex weights which will be an underestimate in general. The final vertex weight will then be formed by combining the two.

For the underestimate, we choose the vertex weights such that only pairs of nearest neighbors will pair with each other.  
Let us define the minimum pairing weight for a vertex $j$,
\be
W^{\min}_j = \min_{k} W_{jk}\,.
\ee
In order to ensure that only mutual nearest neighbors pair with each other, we set the vertex weight to be the `abstaining' weight
\be
W^{A}_j = \frac{W^{\min}_j}{2}+\epsilon\,.
\ee
Here, a small $\epsilon>0$ is used to break the degeneracy between mutual nearest neighbors pairing and both abstaining.

\subsection{Vertex weight}

The tag-along weight is often an overestimate of the ideal vertex weight, and the abstaining weight is an underestimate. A linear interpolation between the two will thus be used:
\be\label{eq:vertexWeight}
W_j = W^{A}_j + \lambda \left(W^T_j - W^{A}_j \right)\,,
\ee
This gives $W^T_j$ at $\lambda=1$ and $W^A_j$ at $\lambda=0$. In general we are free to choose the $\lambda$ for any given $p$, $L$ and $N$ that gives the best compromise between these two methods.
In the following, we set $\lambda=0.3$ throughout, which leads to better results than both $\lambda=0$ and $\lambda=1$. 

However, note that the build-up of degeneracies will sometimes lead $W^T_j$ to become lower than $W^A_j$. This means that $W_j$ becomes smaller than the abstaining weight $W^A_j$, which means that no clusters will pair any more. 
In this case, we resort to the abstaining weight and set $W_j = W^{A}_j$.

Note that in the limit $\lambda\rightarrow0$ the decoder introduced here is similar to the expanding diamonds decoder using shortcuts, in that only mutual nearest neighbors will be fused. However, unlike expanding diamonds, the MWM HDRG decoder can pair mutual nearest neighbors of different distances during the same iteration of the algorithm.

\subsection{Example}

\begin{figure}
\centering
\includegraphics[width=0.9\columnwidth]{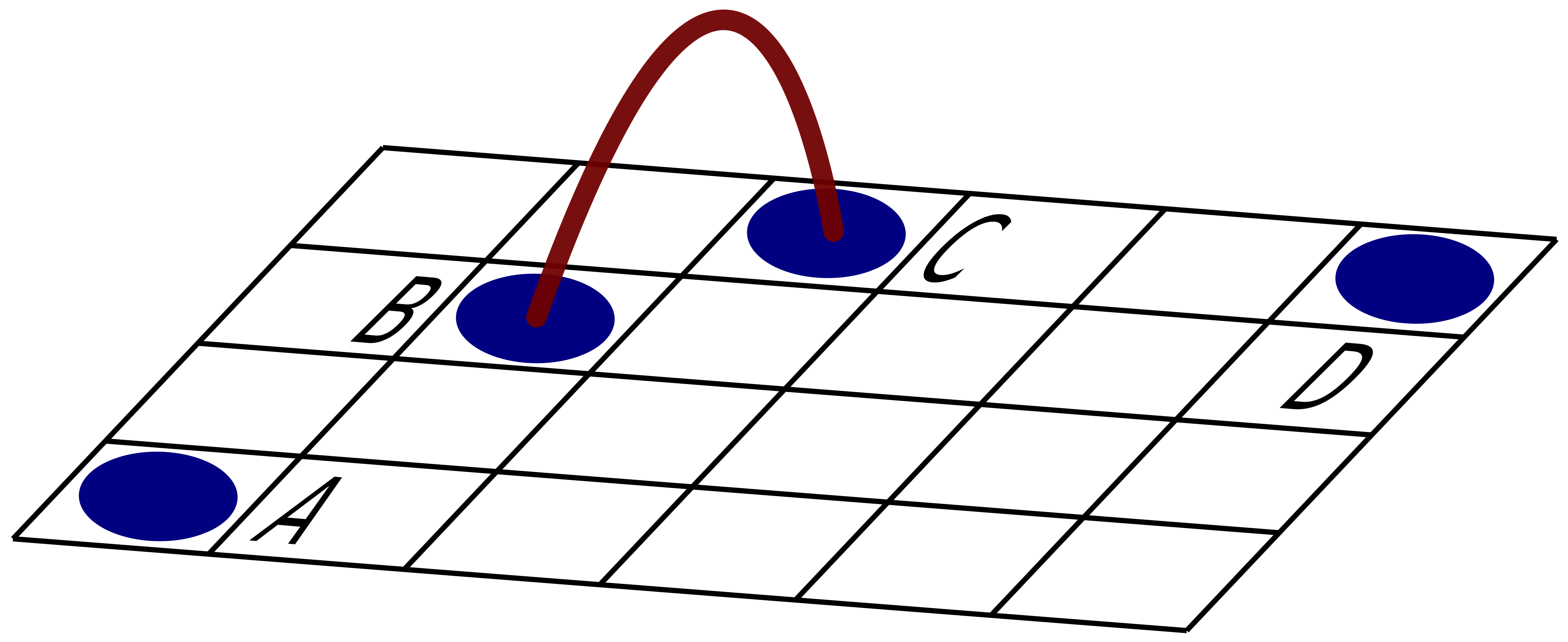}
\caption{An example configuration involving four anyons (blue circles). After fusing anyon $B$ with $C$, we add a ``wormhole'' to the lattice (red arc), which allows other anyons to take shortcuts.}
\label{fig:example}
\end{figure}

Fig.~\ref{fig:example} shows an example configuration of four anyons. Assuming that no fusions have happened so far, we have $W_{AB}=3-\log\binom{3}{1})/\beta$, with $\beta$ as defined in Eq.~(\ref{eq:beta}), $W_A=\frac{1+\lambda}{2}W_{AB}$, etc..
With $d=3$ and $p=10\%$, we have $W_A+W_{BC}+W_D<W_{AB}+W_{CD}$ for $\lambda<0.37$, meaning that the algorithm will fuse anyons $B$ and $C$ in a first round, while anyons $A$ and $D$ refrain from matching at the cost of their vertex weight.
After fusing $B$ with $C$, other anyons are allowed to take shortcuts over the resulting cluster (irrespective of its anyonic charge). This can be thought of as adding a ``wormhole'' to the lattice (the red arc in Fig.~\ref{fig:example}).
Taking the shortcut into account, the weight for connecting anyons $A$ and $D$ is updated from $W_{AD}=8-\log\binom{8}{3}/\beta\approx6.61$ to $W_{AD}=6-\log\binom{2}{1}/\beta\approx5.76$.
For $\lambda>0.37$, anyon $A$ will be paired with anyon $B$ in the first round, as well as $C$ with $D$.

\section{Numerical results for $D(\ZZ_d)$ models}\label{sec:results}

\subsection{Results for perfect syndrome measurements}

\begin{figure}
\centering
\includegraphics[width=0.9\columnwidth]{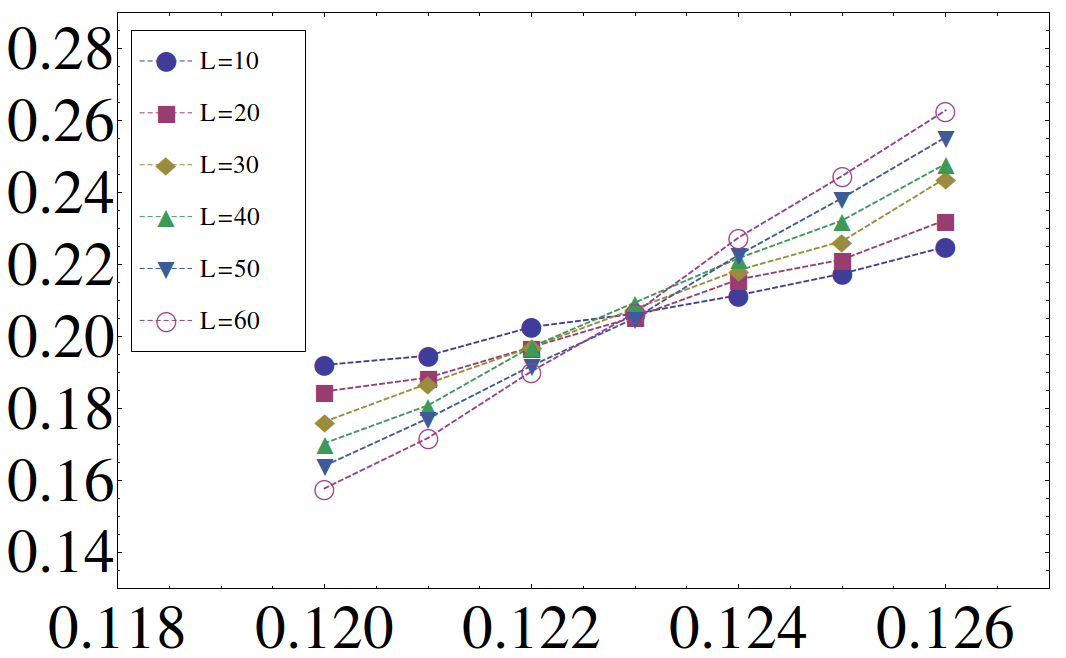}
\caption{Error rate $p$ (horizontal axis) versus logical error rate $p_L$ (vertical axis) for the $D(\ZZ_3)$ model. Each data point represents $10^4$ logical errors.}
\label{fig:Z3PM}
\end{figure}

In this section, we present the results achieved with our MWPM HDRG decoder. Fig.~\ref{fig:Z3PM} shows logical error rates for various values of $p$ and $L$ for the case of $d=3$. We find a cross-over point at $p_c=12.3\%$, indicating the threshold error rate of our decoder.

The cross-over point is obtained from Fig.~\ref{fig:Z3PM} and similar figures by linear interpolation between the logical error rates obtained for equally-sized codes and visual inspection.
More sophisticated fittings are used in Refs.~\cite{harrington_thesis,sdrg,abcb}.

\begin{figure}
\centering
\includegraphics[width=0.9\columnwidth]{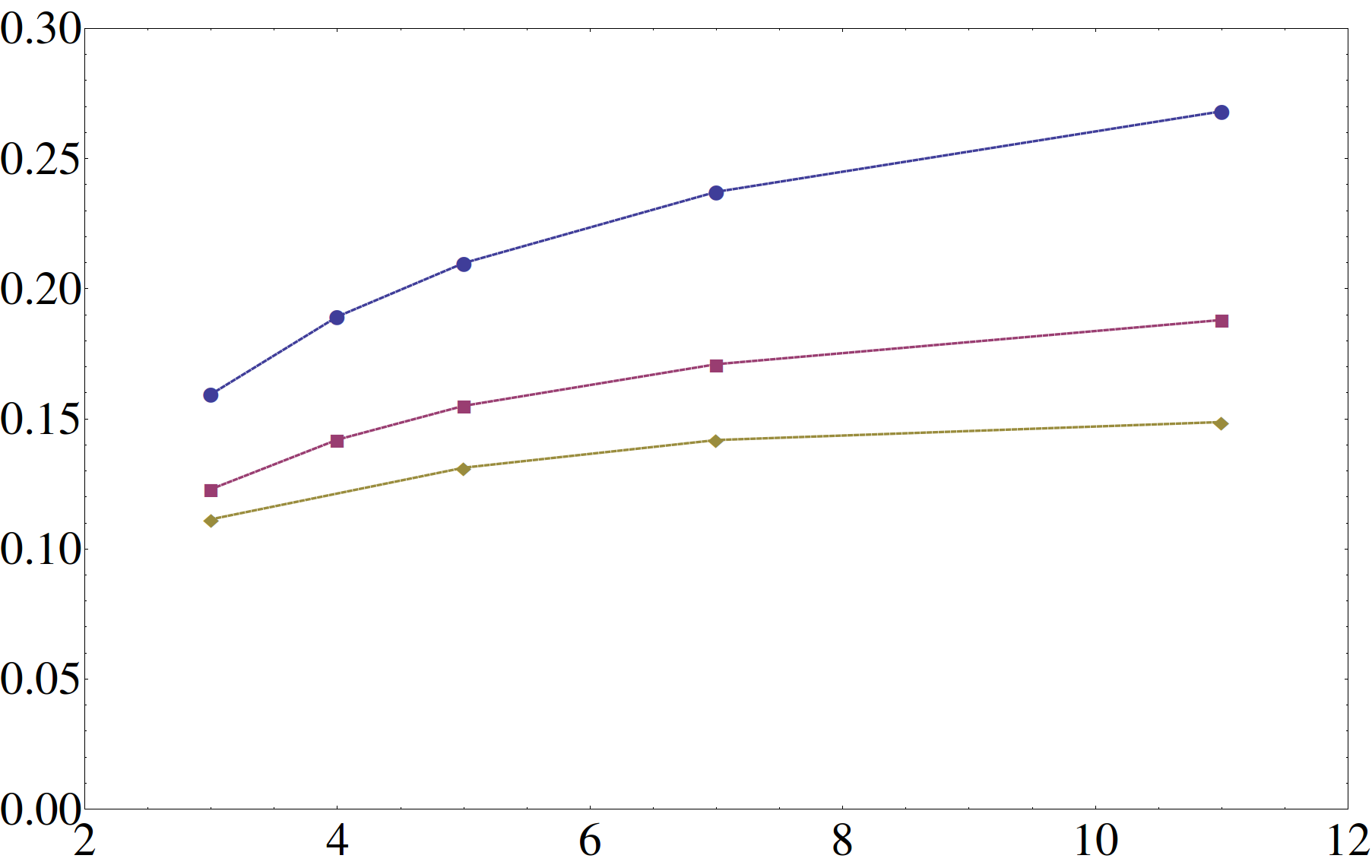}
\caption{Thresholds error rates $p_c$ for the $D(\ZZ_d)$ quantum double models for $d=3,4,5,7,11$.
We show the hashing bound threshold (circles), the threshold achieved with our HDRG decoder (squares), and the threshold achieved by ABCB (diamonds). 
Hashing bound values (circles) are obtained by solving Eq.~(\ref{eq:hash}). 
Our threshold values (squares) are obtained to accuracy $10^{-3}$ by comparing the logical error rates for various values of $p$ and $L=10, 20, \ldots, 60$, as illustrated in Fig.~\ref{fig:Z3PM} for the case $d=3$.}
\label{fig:threshPM}
\end{figure}

We have produced similar plots for low prime dimensions $d=3, 5, 7, 11$ and $d=4$. The corresponding thresholds are displayed in Fig.~\ref{fig:threshPM}. We find these thresholds to be higher than those achieved by HDRG methods in Ref.~\cite{abcb}, yet lower than those achieved with a soft-decision renormalization group (SDRG) decoder in Ref.~\cite{sdrg}.

We also compare our thresholds with the hashing bound threshold, which provides an entropic estimate for the threshold error rates. Indeed, it has recently been shown \cite{andrist} that the maximal threshold error rates for the $D(\ZZ_d)$ models achievable using computationally inefficient methods are very close to the hashing bound threshold values. The hashing bound threshold value for the model $D(\ZZ_d)$ is given by the solution of
\be\label{eq:hash}
-p\log\left(\frac{p}{d-1}\right)-(1-p)\log(1-p) = \frac{1}{2}\log(d)\,.
\ee
The solutions are compared with the threshold values achieved by our algorithm in Fig.~\ref{fig:threshPM}. 

For $d=7919$ (the $1000$-th prime), we find a threshold value of $p_c=21.9\%$, which is significantly above the threshold value for $p$ beyond which the error syndromes start to percolate the code \cite{abcb}.
It is higher than the threshold value achieved by previous HDRG methods \cite{abcb}.

\begin{figure}
\centering
\includegraphics[width=0.9\columnwidth]{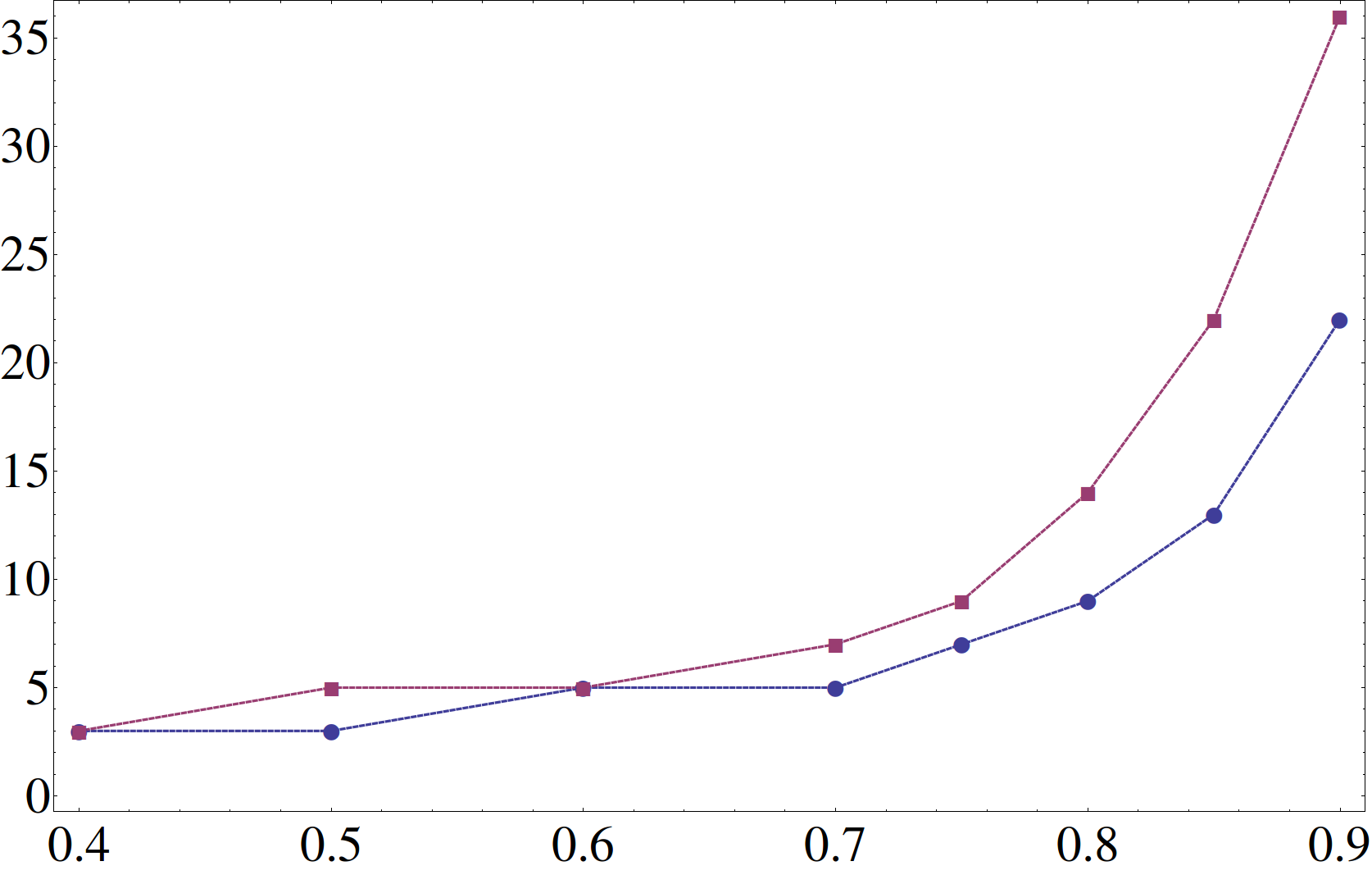}
\caption{
Minimal sizes $L^*(p)$ (vertical axis) such that $p_L<p$ for both $d=3$ (circles) and $d=7919$ (squares) for perfect stabilizer measurements. The horizontal axis shows $p/p_c$ for the threshold values $p_c$ provided in Fig.~\ref{fig:threshPM}. We have $L^*(p)=3$ for all $p/p_c<0.4$.}
\label{fig:LpPM}
\end{figure}

Another important benchmark of a decoder is the minimum system size required such that the logical error rate, $p_L$ is less than the physical error rate, $p$. This value, denoted $L^*(p)$, is the minimum code size for which the error correction yields a positive effect. These sizes were found for extreme cases of $d=3$ and $d=7919$ and are shown in Fig.~\ref{fig:LpPM}. For $p<p_c/2$, system sizes of $L=3$ are sufficient to demonstrate error correction for $d=3$ and $L\leq 5$ is sufficient for $d=7919$. 
Small values of $L^*(p)$ are odd since the minimal number of errors needed to break an $L=2n-1$ code is the same as for an $L=2n$ code. 
At the point of syndrome percolation for $d=7919$, which occurs at around $p=18\%$, a system size of $L=17$ is sufficient to demonstrate error correction.

\subsection{Results for imperfect syndrome measurements}

If syndrome measurements can fail with non-vanishing probability, error correction needs to be performed in a continuous fashion to allow the measurement errors to be detected.
Non-trivial syndromes then persist through time, as long as no (data or syndrome measurement) error happens.
The vertices in the graph entering our HDRG algorithm (which is now three-dimensional) are thus no longer given by non-trivial syndromes, but rather by non-trivial syndrome \emph{changes}.
Fusing two vertices with equal temporal coordinate means presuming data qudit errors, while fusing two vertices with equal spatial coordinates means presuming syndrome measurement errors.

We perform error correction for $L$ rounds and assume that an error-free syndrome measurement is possible after the final round of syndrome measurement.
The same assumption has been made for the qubit case in e.g.\ Ref.~\cite{fowler_fast}.
An alternative would be to assume periodic boundary conditions in temporal direction \cite{sdrg_continuous}.
While both of these assumptions cannot be justified on physical grounds, they are necessary in order to observe a threshold error rate without explicitly modelling a measurement of the non-locally stored quantum information.
In reality, the logical quantum state would have to be measured in a fault-tolerant way, and we avoid explicit modelling of such a measurement process for simplicity.

\begin{figure}
\centering
\setlength{\unitlength}{0.9\columnwidth}
  \begin{picture}(1.0,0.65)
	\put(0.02,0.0){\includegraphics[width=\unitlength]{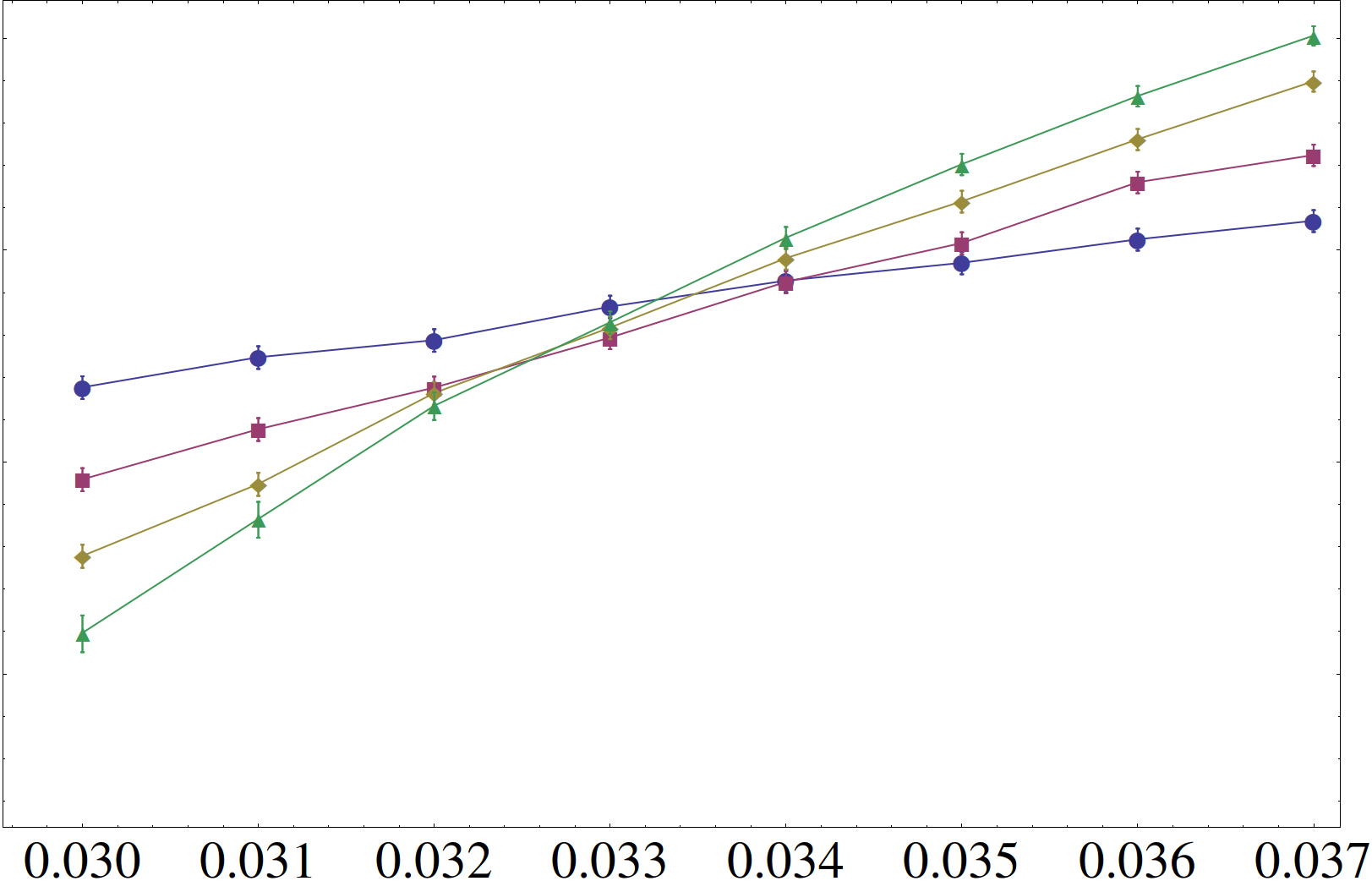}}
	\put(-0.058,0.455){$0.10$}
	\put(-0.058,0.146){$0.01$}
  \end{picture}
\caption{Error rate $p$ (horizontal axis) versus logical error rate $p_L$ (vertical axis) for $L=8, 16, 24, 32$ (from top to bottom at $p=0.030$) for the $D(\ZZ_3)$ model. 
Each data point represents $10^3$ logical errors, or at least $400$ for $L=32$ and low $p$.
Error bars are taken to be $2\sigma$.
We notice considerable finite-size effects for $L=8$.}
\label{fig:Z3imperfect}
\end{figure}

We model syndrome measurement errors by adding with probability $p$ one of the $d-1$ non-trivial values $1, \ldots, d-1$ to the actual syndrome value (modulo $d$).
This generalizes the modelling of syndrome measurement errors used for the qubit case in Refs.~\cite{harrington_thesis,sdrg_continuous}.
The distance between two non-trivial syndrome changes is then given by the 3D Manhattan distance $d_{jk}=d_{jk}^x+d_{jk}^y+d_{jk}^t$, and the number of possible minimum-weight error paths connecting them is
\be
\mu_{jk} = (d-1) {d_{jk} \choose d^t_{jk}} {d_{jk}^x+d_{jk}^y \choose d^x_{jk}}\,.
\ee

Since the logical errors in our Monte Carlo simulations follow a binomial distribution, the standard deviation in the logical error rates are given by $\sigma=\sqrt{p_L(1-p_L)/N}$, where $N$ is the number of experiments.
Fig.~\ref{fig:Z3imperfect} shows $2\sigma$ error bars.
From Fig.~\ref{fig:Z3imperfect}, we estimate a threshold value of $3.2\%$ for the $d=3$ case. 
This is larger than the thresholds obtained with an analogous error model for the qubit ($d=2$) case. 
Minimum-weight perfect matching achieves in this case a threshold of $2.9\%$ \cite{harrington_thesis}, while an SDRG decoder achieves $1.9\%$ \cite{sdrg_continuous}. 

\begin{figure}
\centering
\includegraphics[width=0.9\columnwidth]{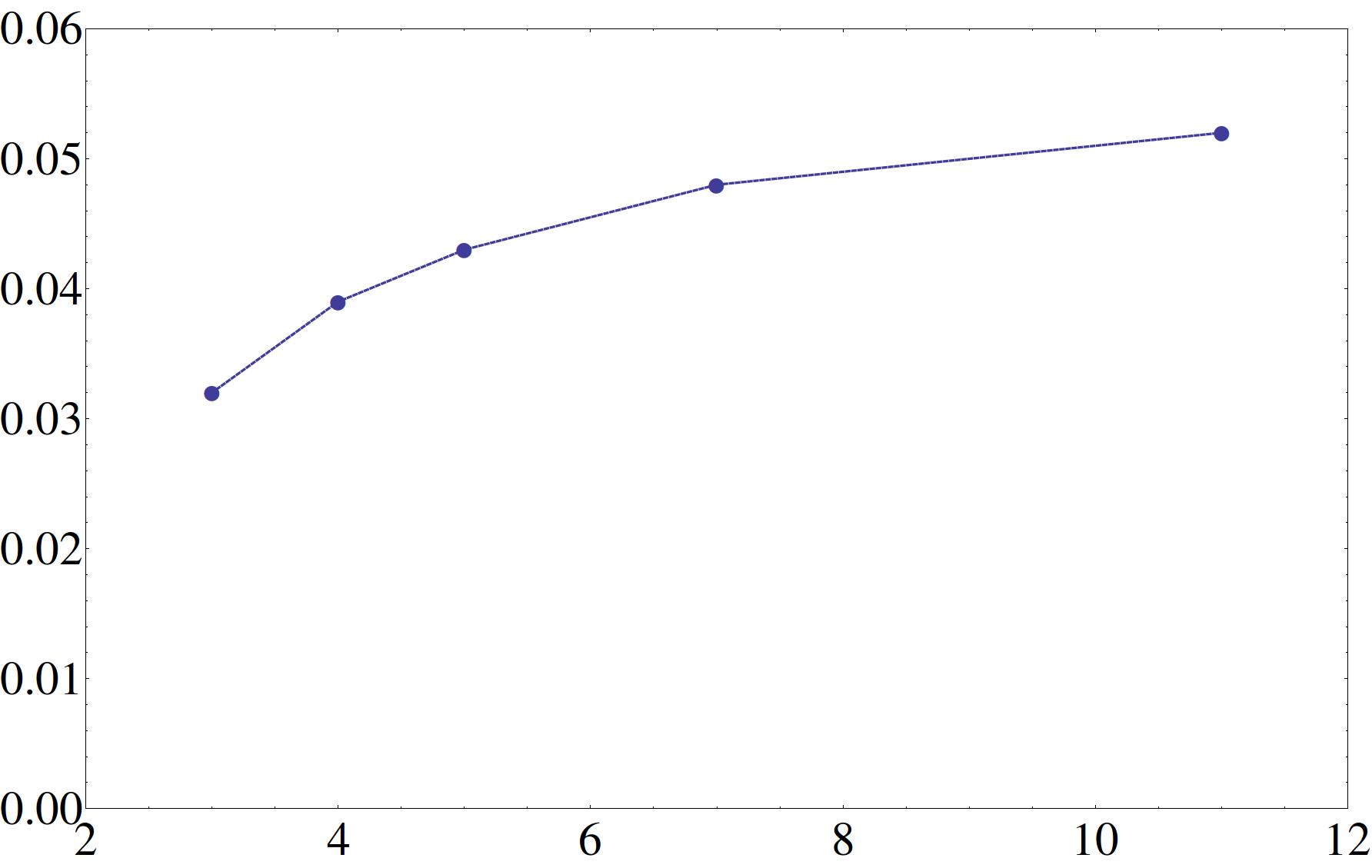}
\caption{Thresholds $p_c$ achieved with our HDRG decoder for $d=3,4,5,7,11$ when errors affect both data qudits and syndrome measurements with a rate $p$.}
\label{fig:threshImperfect}
\end{figure}

Finally, Fig.~\ref{fig:threshImperfect} shows the thresholds obtained by comparing logical error rates as in Fig.~\ref{fig:Z3imperfect} for different values of $d$.
Note that for the imperfect measurement case, there is no obvious generalization of the Hashing bound with which our thresholds could be compared.
For $d=7919$, we obtain a threshold of $p_c=6.1\%$.

\begin{figure}
\centering
\includegraphics[width=0.9\columnwidth]{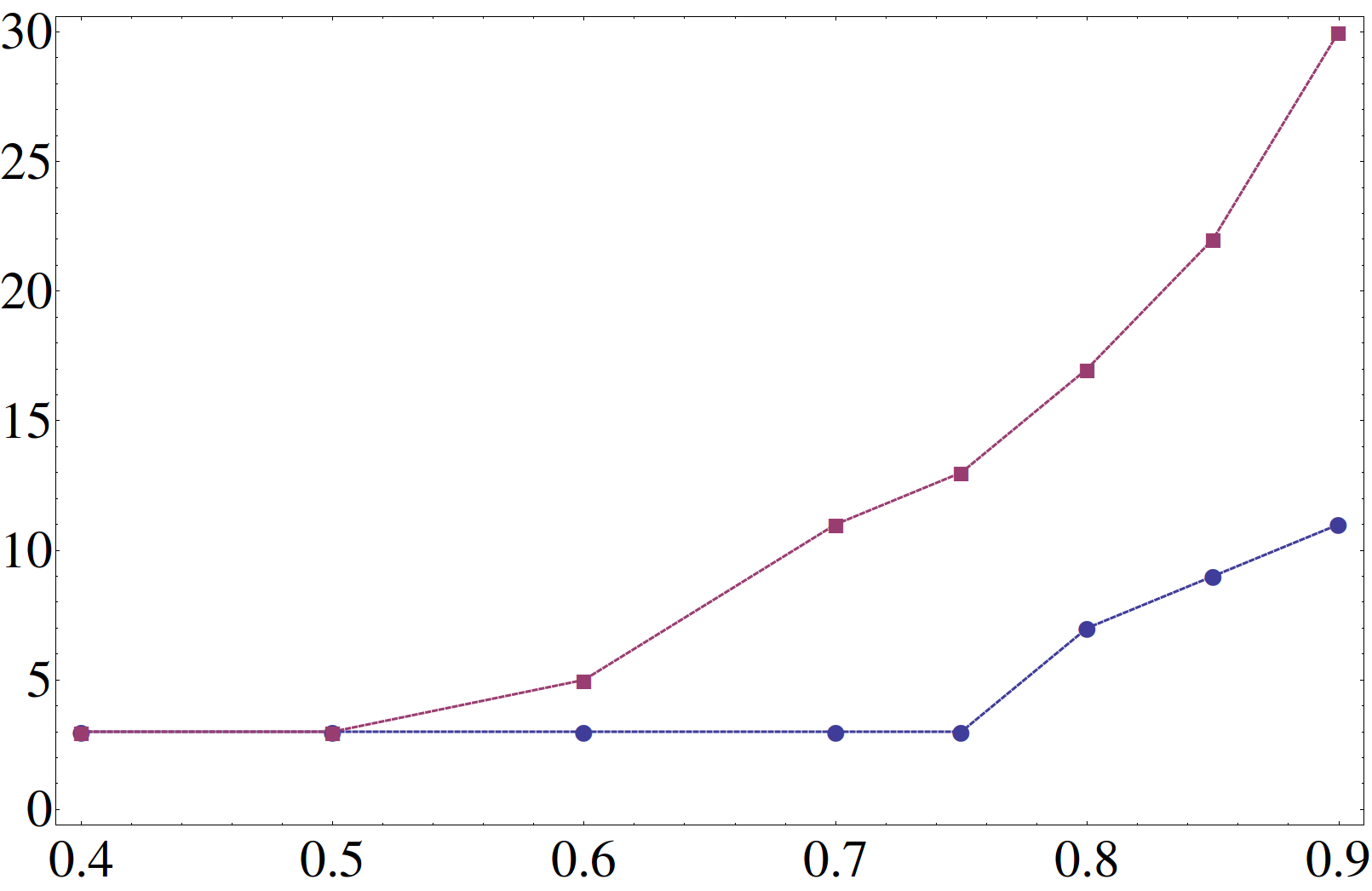}
\caption{
Minimal sizes $L^*(p)$ (vertical axis) such that $p_L<p$ for both $d=3$ (circles) and $d=7919$ (squares) for imperfect stabilizer measurements. The horizontal axis shows $p/p_c$ for the threshold values $p_c$ provided in Fig.~\ref{fig:threshPM}. We have $L^*(p)=3$ for all $p/p_c<0.4$.}
\label{fig:LpIPM}
\end{figure}

We have again determined the minimal code sizes $L^*(p)$ which are necessary to achieve $p_L<p$ for some $p$ for $d=3$ and $d=7919$. The results are given as a function of $p/p_c$ in Fig.~\ref{fig:LpIPM}.

\section{Decoding Non-Abelian anyons}\label{sec:nonabelian}

Due to the way HDRG decoders have been defined in this work, they are directly applicable to the case of non-Abelian anyons. This can be demonstrated by using them to decode the $\Phi-\Lambda$ model, a non-Abelian model whose anyons have fusion behaviour similar to that of the Fibonacci model \cite{wootton_na}. Specifically, they have the fusion rules
\be
\Lambda \times \Lambda = 1\,, \,\,\, \Lambda \times \Phi = \Phi\,, \,\,\, \Phi \times \Phi = 1 + \Lambda + \Phi\,.
\ee
Note that the $\Phi$ and $\Lambda$ anyons are their own antiparticles.

Except for the fusion channel to a $\Phi$ in the last fusion rule, the fusion rules of the $\Phi-\Lambda$ model are identical to those for Ising anyons:
\be
\psi \times \psi = 1\,, \,\,\, \psi \times \sigma = \sigma\,, \,\,\, \sigma \times \sigma = 1 + \psi\,.
\ee
Decoding of this model was studied both with the BH decoder and using MWPM methods in Ref.~\cite{brell_na}. In order to understand decoding by means of MWPM, note that $\sigma$ anyons can only be created and destroyed in pairs. It is thus possible to temporarily treat $\psi$ particles as vacuum and use MWPM to pair all $\sigma$ particles. In a second round, MWPM can be used to pair all $\psi$ particles.

Similarly, it is possible to decode the $\Phi-\Lambda$ model by first fusing all $\Phi$ anyons and then pairing all remaining $\Lambda$ particles by use of MWPM. However, in contrast to the Ising model, we can no longer use MWPM for the first round of decoding. Two $\Phi$ anyons can fuse both to a non-$\Phi$ outcome ($1$ or $\Lambda$) or to another $\Phi$ particle, exhibiting Fibonacci-like behavior. (In particular, the number of $\Phi$ anyons need not be even, as required for MWPM.) It is thus necessary to apply HDRG methods for this first round of decoding.

We consider the case of non-Abelian decoding with perfect syndrome measurements. In this case the $\Phi-\Lambda$ model can be efficiently simulated by the Abelian $D(\ZZ_6)$ model \cite{wootton_na}. 
A $\Lambda$ thereby corresponds to a charge $m_3$, while a $\Phi$ corresponds to charges $m_1$, $m_2$, $m_4$, and $m_5$.
The simulation requires that the decoder cannot distinguish between the different charges of the $D(\ZZ_6)$ model that correspond to a $\Phi$.
Any more information would correspond to the decoding accessing the internal fusion space of the $\Phi$ anyons in an illegal way, and so no longer provides a good simulation of the non-Abelian model. 

When applying the MWPM algorithm for pairing the $\Lambda$ particles, the pairing weights between two of them would ideally incorporate knowledge about the initial location of all $\Phi$ anyons that fused into a particular $\Lambda$.
However, for simplicity we ignore knowledge about the fusion history of the $\Lambda$ particles during MWPM.

\begin{figure}
\centering
\includegraphics[width=0.9\columnwidth]{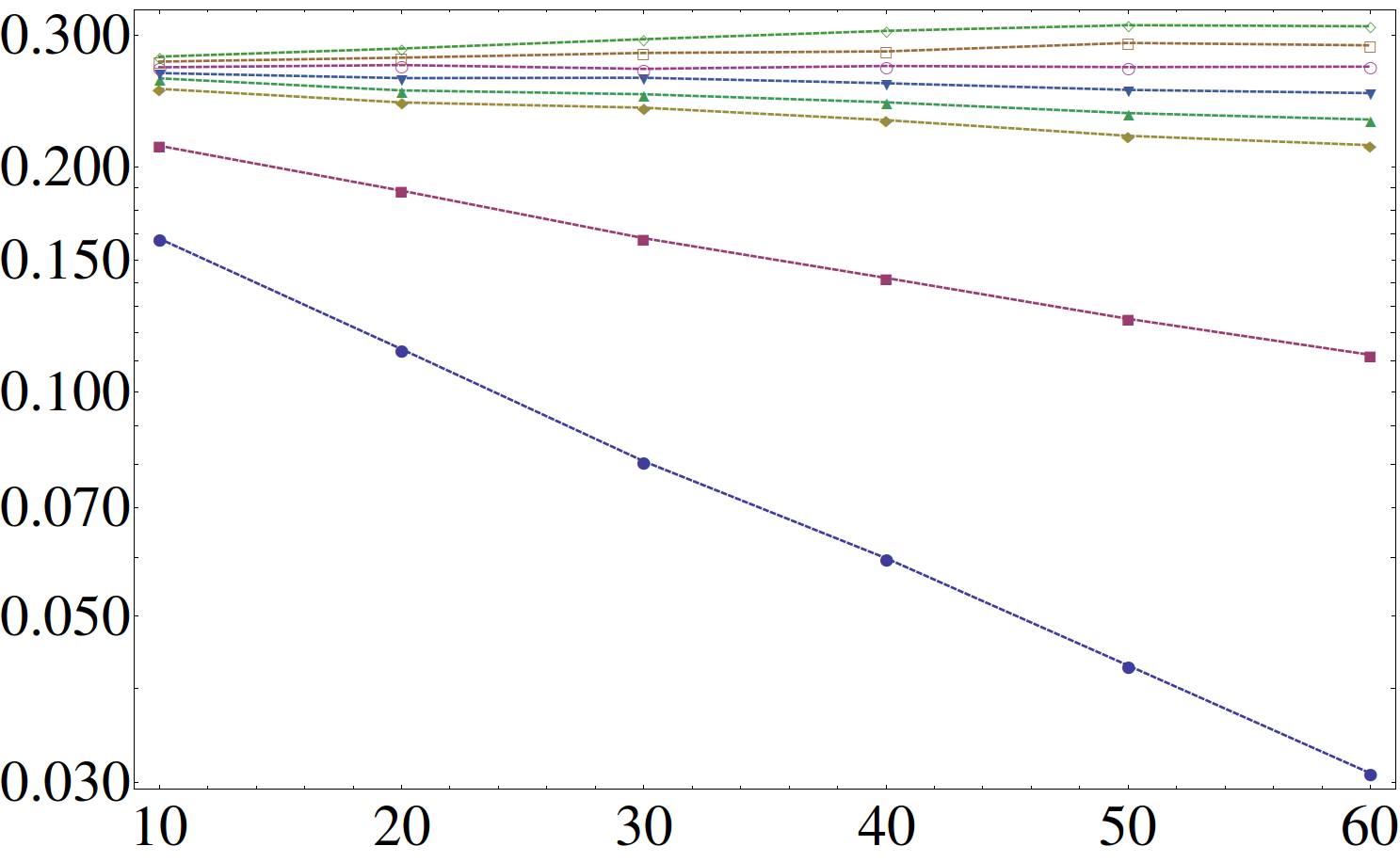}
\caption{Logical error rate $p_L$ as a function of $L$ for various error rates $p$. From top to bottom, we have $p=0.152, 0.151, 0.150, 0.149, 0.148, 0.147, 0.140, 0.130$. A threshold at $p_c=15.0\%$ and exponential suppression of $p_L$ for $p<p_c$ are clearly recognizable. Data points represent $10^4$ logical errors.}
\label{fig:nonabTresh1}
\end{figure}

\begin{figure}
\centering
\includegraphics[width=0.9\columnwidth]{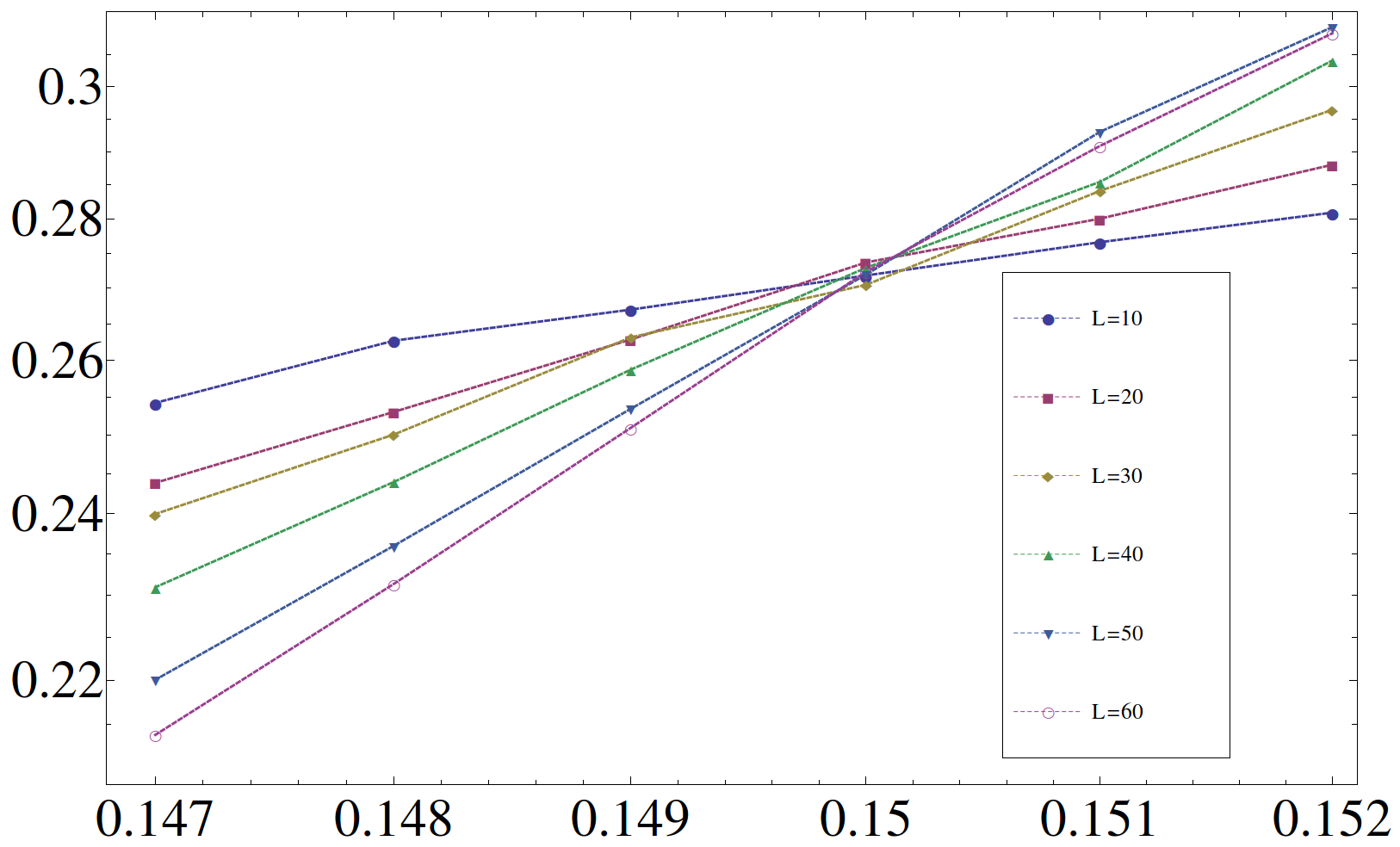}
\caption{Logical error rate $p_L$ as a function of $p$ close to the threshold for various $L$.}
\label{fig:nonabTresh2}
\end{figure}

We consider an error model in which $p_{\Phi} = p_{\Lambda} = p/2$. In terms of the $D(\ZZ_6)$ model used for the simulation, we have  $p_1 = p_2 = p_4 = p_5 = p_\Phi/4$, while $p_3 = p_\Lambda$. Here, $p_g$ denotes the probability of a $(\sigma^x)^{g}$ error in the $D(\ZZ_6)$ model. Ref.~\cite{wootton_na} employed the expanding diamonds decoder for this error model and found a threshold error rate of $p_c=7.0\%$. Figs.~\ref{fig:nonabTresh1} and \ref{fig:nonabTresh2} show that our decoder achieves a threshold of $p_c=15.0\%$, more than twice as high as the one achieved by previous HDRG methods.
Fig.~\ref{fig:nonabTresh1} suggests a scaling of the form $p_L\sim\exp[-\alpha(p)L^1]$. Recall from our discussion in Sec.~\ref{sec:cantor} that this improvement over the $p_L\sim\exp[-\alpha(p)L^{2/3}]$ scaling achieved by previous HDRG decoders is due to the use of shortcuts. We point out again that even when using the shortcuts, there will be sub-polynomial corrections to the linear-in-$L$ exponent.

\begin{figure}
\centering
\includegraphics[width=0.9\columnwidth]{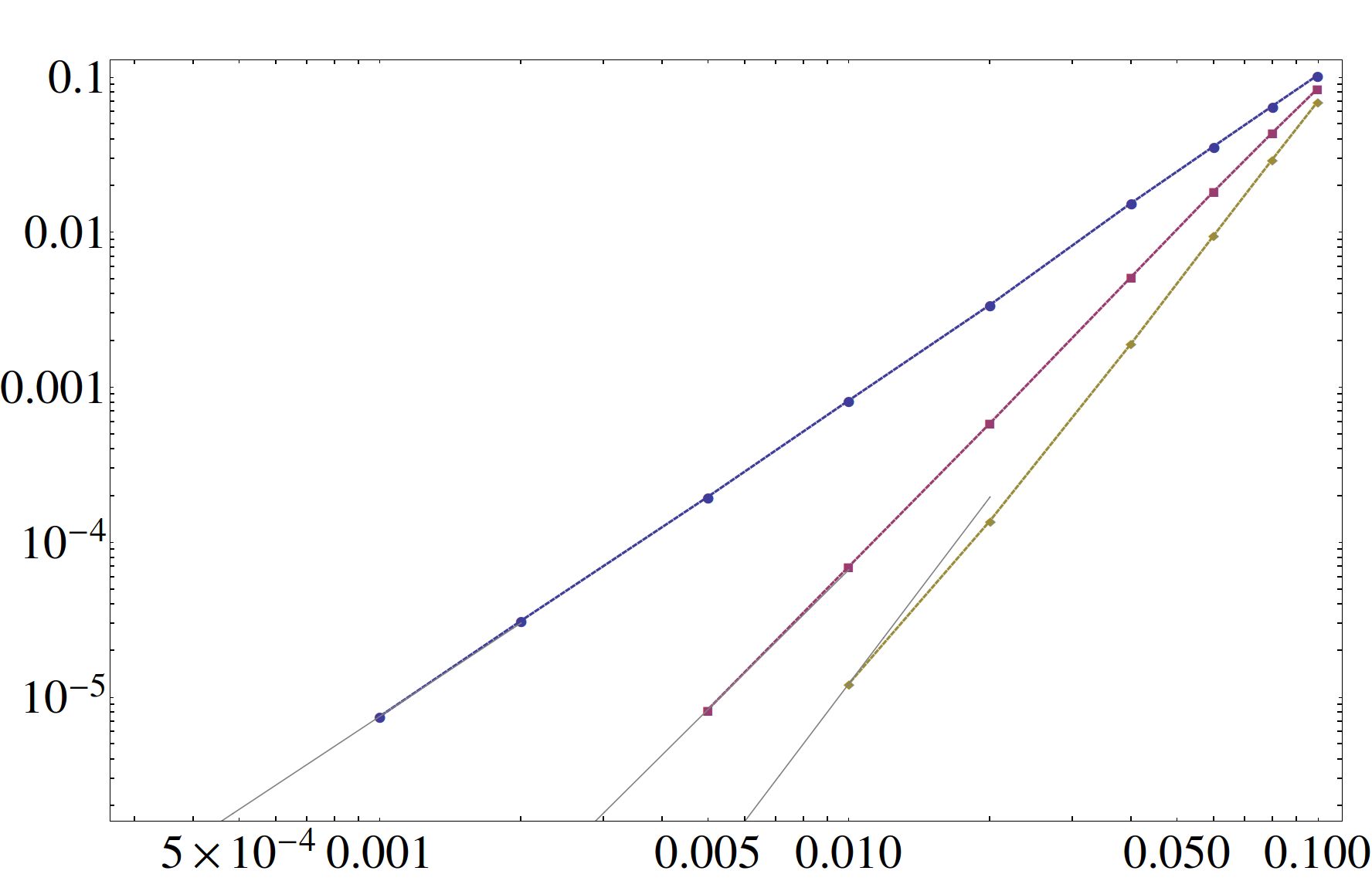}
\caption{Logical error rate $p_L$ as a function of $p$ for small-distance codes $L=3,5,7$ (top to bottom). Gray lines are fittings of the form $ap^{(L+1)/2}$ through the lowest data point for each $L$, showing that for $L=3,5$ we are already well in the regime where most likely error chains dominate the logical error rate.}
\label{fig:nonabLowL}
\end{figure}

Fig.~\ref{fig:nonabLowL} provides logical error rates in the low-$p$, low-$L$ regime and shows that our decoding indeed allows the code to use its whole distance.
Recall that the use of shortcuts makes $\lfloor(L+1)/2\rfloor$ errors necessary for a logical error for $L<9$, leading to a suppression $p_L\sim p^{\lfloor(L+1)/2\rfloor}$ for low enough $p$.

The case of imperfect syndrome measurements for non-Abelian anyons is more complex than Abelian ones, and so cannot be done simply through the case of noisy syndrome measurements in $D(\ZZ_6)$. This will be addressed in future work.

\section{Runtime of our algorithm}\label{sec:runtime}

In this section, we provide a heuristic estimate of the parallelized runtime of our algorithm, for both the case with perfect and imperfect syndrome measurements.

\begin{figure}
\setlength{\unitlength}{0.9\columnwidth}
  \begin{picture}(1.0,0.65)
	\put(-0.05,0.0){\includegraphics[width=1.1\unitlength]{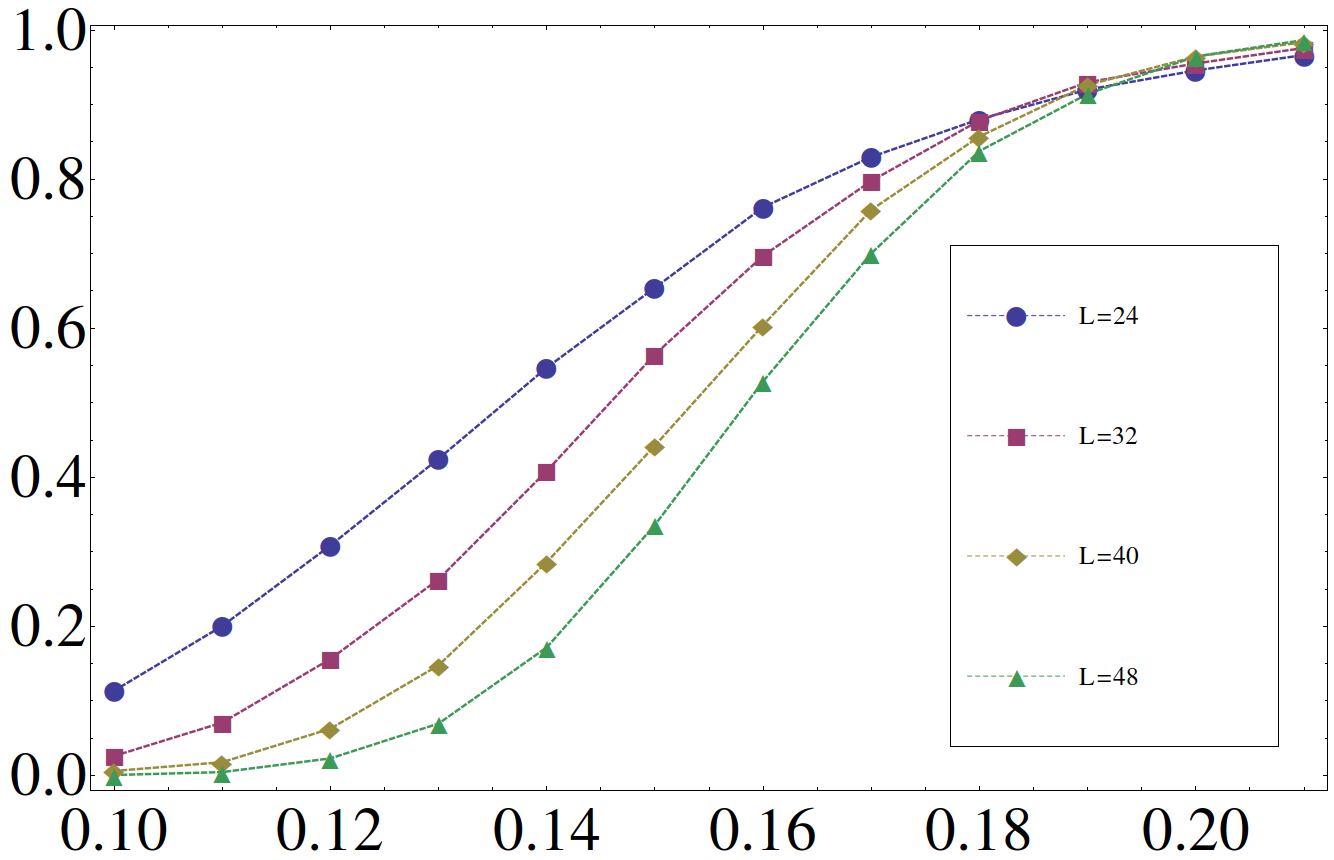}}
	\put(0.04,0.43){\includegraphics[width=0.38\unitlength]{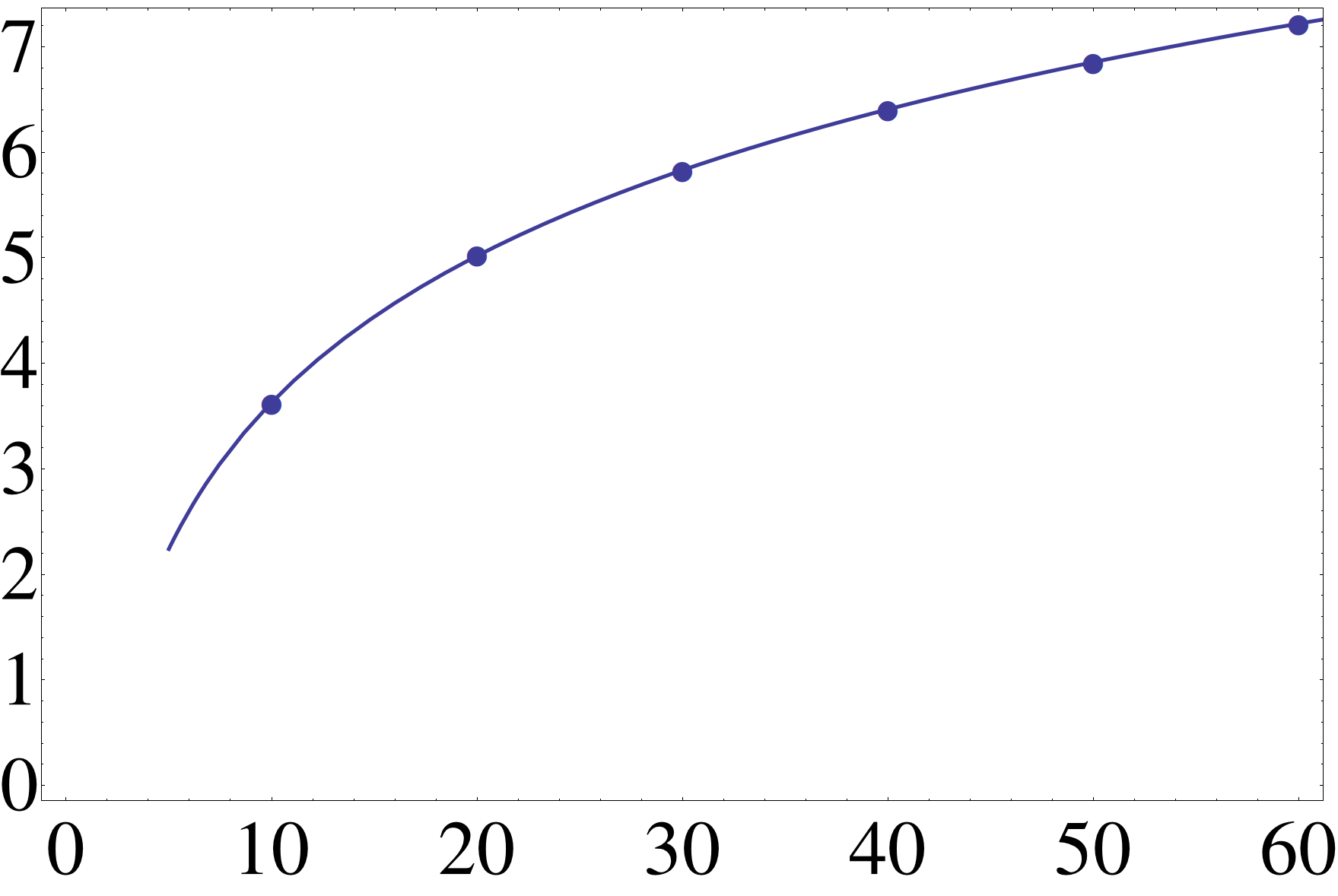}}
  \end{picture}
\caption{Propability that a subgraph wraps around the entire code for various error rates $p$ (horizontal axis) and code sizes $L$ for the $D(\ZZ_3)$ case with perfect measurements.
Two vertices (non-trivial syndrome measurements) are connected by an edge if their distance is strictly smaller than the sum of their nearest-neighbor distances. The Manhattan distance is used for simplicity.
A crossover point is observed at roughly $p=19\%$, below which the probability of a code-spanning subgraph vanishes as $L\rightarrow\infty$. 
The inset shows the average number of iterations of our algorithm necessary for $p=12\%$ as a function of $L$. The line is a fit of the form $a\log L + b$.}
\label{fig:percolation}
\end{figure}

Recall that in our algorithm each vertex has the possibility to ``self-match" at the cost of the vertex-weight given in Eq.~(\ref{eq:vertexWeight}), which is upper-bounded by the Manhattan distance to its nearest neighbor.
Two vertices will thus only ever be matched by the algorithm if their distance is smaller than the sum of their respective nearest-neighbor distances.
If their distance is larger, it is thus unnecessary to add an edge between them.
For low enough $p$, the typical nearest-neighbor distance is $O(1)$ (an anyon can only be created from the anyonic vacuum together with another anyon), while the typical next-to-nearest-neighbor distance is $O(p^{-1/2})$.
Each vertex is thus typically only connect to one other vertex for low enough $p$.
This means that the graph given to the perfect matching algorithm decays into subgraphs of average size $O(1)$.
The threshold value above which one of the subgraphs obtained this way percolates the entire code is estimated for the $D(\ZZ_3)$ case with perfect measurements in Fig.~\ref{fig:percolation}. It is significantly higher than the threshold error rate of our algorithm.
Our algorithm thus lends itself nicely to parallelization. 
Note that the shortcuts discussed in Sec.~\ref{sec:improving} lead to local deformations of the lattice geometry only.

If $p$ is below the aforementioned threshold, the propability of a subgraph involving $n$ vertices is exponentially small in $n$.
Correspondingly, the maximal number of vertices we expect to find in a subgraph is for a code of linear size $L$ given by $O(\log L)$, as is well-known from percolation theory.
For a graph with $n$ vertices and $O(n^2)$ edges, the perfect matching algorithm \texttt{Blossom V} \cite{blossom} finds a MWPM in time $O(n^3\log n)$.
In conclusion, one iteration of our MWPM HDRG algorithm takes in the perfect measurement case a time which grows like $\text{poly}(\log L)$.

The lower $\lambda$ in the vertex-weight Eq.~(\ref{eq:vertexWeight}) is, the cheaper it is for an anyon to self-match and refrain from fusing with another anyon. The smallest number of fusions occurs for $\lambda=0$, where two anyons are only fused if they are mutual nearest neighbors. Since we expect the number of mutual nearest neighbor pairs among all anyons not to fall below a certain fraction, at least a certain fraction of anyons will fuse during each iteration of the algorithm, such that $O(\log L)$ iterations will be sufficient even for $\lambda=0$.
The inset of Fig.~\ref{fig:percolation} shows the average number of iterations of our algorithm for $d=3$ and $\lambda=0.3$, clearly following a logarithmic trend.
With an average of $O(\log L)$ iterations, the total expected runtime of our algorithm is $\text{poly}(\log L)$.

For the more realistic case with imperfect measurements, where error correction is performed in a continuous fashion, the relevant quantity is the classical processing time per round of error correction.
We assume that the error rate $p$ is the same for data qubit errors and for syndrome measurement errors, and that we perform error correction for $L$ time-steps.
After including measurement errors, three-dimensional clusters of syndrome changes will still be of average size $O(1)$ and maximal size $O(\log L)$.
If the \emph{local} processing speed of the classical computing devices performing the error correction algorithm can be temporarily increased by a factor of $2$, 
larger than average sized clusters can still be dealt with in constant average time, as they are exponentially unlikely.
Such an approach to error correction with constant average processing time per round of error correction has been described in much more detail in Ref.~\cite{fowler_proof}.

\section{Conclusions}\label{sec:conclusions}

In conclusion, we have discussed strengths and weaknesses of existing HDRG decoders, and have proposed a new minimum-weight matching based algorithm which does not force us to compromise between the advantages of the different  algorithms. Indeed, we have shown that in the perfect measurement case for the $D(\ZZ_d)$ quantum double models our algorithm achieves higher thresholds than previous HDRG decoders. Furthermore, we have used it to perform the first study of error correction for these qudit topological codes for which the possibility of syndrome measurement failure is taken into account. 

The defining feature of non-Abelian systems is that the outcome of fusing two defects cannot be predicted when given local properties of the two defects only. The information about the fusion outcome is stored in non-local degrees of freedom, which are used to store and process quantum information. Since our decoder uses only the geometrical location of defects as inputs, and then updates based on whether or not two defects can be brought to annihilation, the methods discussed in this work are straightforwardly applicable to non-Abelian systems.
We have employed them to achieve a drastically increased error threshold for a particular non-Abelian model, and anticipate their application in the open problem of continuous error correction for non-Abelian systems.

\section{Acknowledgements}

The authors would like to thank Benjamin Brown for critical reading of the manuscript and sharing data, and the Swiss NF and NCCR QSIT for support

\emph{Note.} While this work was in preparation, the authors learnt of other forthcoming results for noisy syndrome measurements on the qudit codes \cite{watson}. This provides non-HDRG methods that could be used in conjunction with our decoder to boost performance.


\end{document}